\documentclass[fleqn,usenatbib]{mnras}

\usepackage[T1]{fontenc}
\usepackage{ae,aecompl}
\usepackage{soul}
\usepackage{courier}

\usepackage{graphicx}	
\usepackage{amsmath}	
\usepackage{amssymb}	
\usepackage{subcaption}
\title[15\,GHz interday variability of blazars]{The presence of interstellar scintillation in the 15\,GHz interday variability of 1158 OVRO-monitored blazars}

\author[J. Y. Koay et al.]{J. Y. Koay,$^{1}$\thanks{E-mail: jykoay@asiaa.sinica.edu.tw}
D. L. Jauncey,$^{2,3}$
T. Hovatta,$^{4,5}$
S. Kiehlmann,$^{6,7,8}$
H. E. Bignall,$^{9}$
\newauthor W. Max-Moerbeck,$^{10}$
T. J. Pearson,$^{6}$
A. C. S. Readhead,$^{6}$
R. Reeves,$^{11}$
\newauthor C. Reynolds,$^{9}$
H. Vedantham$^{12,13}$
\\
\\
$^{1}$Institute of Astronomy and Astrophysics, Academia Sinica, Section 4, Roosevelt Rd., Taipei 10617, Taiwan\\
$^{2}$CSIRO Astronomy and Space Science, Epping 1710, Australia\\
$^{3}$Research School of Astronomy and Astrophysics, Australian National University, Canberra 2611, Australia\\
$^{4}$Finnish Centre for Astronomy with ESO (FINCA), University of Turku, FI-20014, Turku, Finland \\
$^{5}$Aalto University Mets\"{a}hovi Radio Observatory, Mets\"{a}hovintie 114, 02540 Kylm\"{a}l\"{a}, Finland\\
$^{6}$Owens Valley Radio Observatory, California Institute of Technology, Pasadena, CA 91125, USA\\
$^{7}$Institute of Astrophysics, Foundation for Research and Technology-Hellas, GR-71110 Heraklion,Greece\\
$^{8}$Department of Physics, University of Crete, GR-70013 Heraklion, Greece\\
$^{9}$CSIRO Astronomy and Space Science, Kensington 6151, Australia\\
$^{10}$Departamento de Astronom\'{\i}a, Universidad de Chile, Camino El Observatorio 1515, Las Condes, Santiago, Chile \\
$^{11}$Departamento de Astronom\'{\i}a, Universidad de Concepti\'{o}n, Concepci\'{o}n, Chile\\
$^{12}$Cahill Center for Astronomy and Astrophysics, California Institute of Technology, 1200 E. California Blvd. Pasadena, CA 91125, USA\\
$^{13}$Netherlands Institute for Radio Astronomy (ASTRON), Oude Hogeveensedijk 4, NL-7991 PD Dwingeloo, the Netherlands\\
}

\date{Accepted XXX. Received YYY; in original form ZZZ}

\pubyear{2017}

\begin{document}
\label{firstpage}
\pagerange{\pageref{firstpage}--\pageref{lastpage}}
\maketitle

\begin{abstract}
We have conducted the first systematic search for interday variability in a large sample of extragalactic radio sources at 15\,GHz. From the sample of 1158 radio-selected blazars monitored over a $\sim$10 year span by the Owens Valley Radio Observatory 40-m telescope, we identified 20 sources exhibiting significant flux density variations on 4-day timescales. The sky distribution of the variable sources is strongly dependent on the line-of-sight Galactic H$\alpha$ intensities from the Wisconsin H$\alpha$ Mapper Survey, demonstrating the contribution of interstellar scintillation (ISS) to their interday variability. 21\% of sources observed through sight-lines with H$\alpha$ intensities larger than 10\,rayleighs exhibit significant ISS persistent over the $\sim$10 year period. The fraction of scintillators is potentially larger when considering less significant variables missed by our selection criteria, due to ISS intermittency. This study demonstrates that ISS is still important at 15\,GHz, particularly through strongly scattered sight-lines of the Galaxy. Of the 20 most significant variables, 11 are observed through the Orion-Eridanus superbubble, photoionized by hot stars of the Orion OB1 association. The high-energy neutrino source TXS\,0506$+$056 is observed through this region, so ISS must be considered in any interpretation of its short-term radio variability. J0616$-$1041 appears to exhibit large $\sim$20\% interday flux density variations, comparable in magnitude to that of the very rare class of extreme, intrahour scintillators that includes PKS0405$-$385, J1819$+$3845 and PKS1257$-$326; this needs to be confirmed by higher cadence follow-up observations.
\end{abstract}




\section{Introduction}\label{intro}

The radio variability of compact Active Galactic Nuclei (AGNs) provides a probe of extreme jet physics on scales comparable to or even exceeding that probed using VLBI techniques. Based on light-travel time arguments, variations observed on the shortest timescales are expected to originate from the most compact regions, although this is complicated by the effects of relativistic beaming in blazars.

A further complication arises from interstellar scintillation \citep[ISS, ][]{heeschenrickett87,rickett90,jaunceyetal00}, which has been shown to dominate blazar variability on timescales of a few days or less at cm wavelengths. The 5\,GHz Micro-Arcsecond Scintillation-Induced Variability (MASIV) Survey \citep{lovelletal08} found that $\sim$60\% of 500 compact flat-spectrum AGNs monitored exhibit 2 to 10\% flux density variations on 2-day timescales due to ISS. A follow-up survey \citep{koayetal11} also found ISS to dominate the intra and interday flux density variations at 8\,GHz, as seen in other scintillation studies \citep[e.g., ][]{rickettetal06}.

While ISS has been observed in individual sources at 15\,GHz \citep[e.g., ][]{savolainenkovalev08}, there are no similar large-scale statistical studies of ISS at 15\,GHz; variability at these frequencies is typically assumed to be predominantly intrinsic to the sources themselves.  

The Owens Valley Radio Observatory (OVRO) blazar monitoring program \citep{richardsetal11} provides a rich dataset for studying AGN variability at 15\,GHz. It is the largest and most sensitive radio monitoring survey of blazars, and has been ongoing since the year 2008. The full sample of this OVRO monitoring program now comprises $\sim 1830$ sources, each observed at a cadence of about twice a week, barring bad weather conditions and hardware issues. 

 The OVRO data have been used extensively to estimate the variability brightness temperatures of blazars \citep[e.g.,][]{liodakisetal18a}, study their radio-gamma ray relationship \citep[e.g.,][]{max-moerbecketal14,richardsetal14} and perform multi-frequency cross-correlation studies of blazar flares \citep[e.g.,][]{hovattaetal15,liodakisetal18b,pushkarevetal19}. In these studies, the 15\,GHz flux density variations are always assumed to be intrinsic to the blazar jets. Indeed, the source variability amplitudes from the OVRO lightcurves, as quantified by the intrinsic modulation index \citep{richardsetal11}, broadly show no significant Galactic dependence \citep{koayetal18}, confirming that intrinsic variations likely dominate. This is to be expected since this method of variability characterization is biased towards the largest inflections observed at the longest timescales in the lightcurves, most of which are expected to be intrinsic to the blazars.

The only major studies of interstellar scattering using data from the OVRO monitoring program involved the sources J2025$+$3343 \citep{karaetal12,pushkarevetal13} and J1415$+$1320 \citep{vedanthametal17a}. Symmetric U-shaped features observed in their lightcurves were attributed to or modelled as extreme scattering events \citep[ESEs,][]{fiedleretal87}, arising from lensing by high-pressure intervening clouds of unknown origin in the interstellar medium. ESEs were subsequently ruled out as an explanation for J1415$+$1320 due to the achromatic behavior of the U-shaped features up to mm-wavelengths \citep{vedanthametal17a,vedanthametal17b}; the variations are instead ascribed to gravitational lensing by intervening structures. 

Some questions remain -- Is there significant variability in the OVRO blazar lightcurves on the shortest observed interday timescales? If so, are these interday flux density variations intrinsic to the AGN or due to ISS? How prevalent is ISS at 15\,GHz? Answering these questions is crucial for the interpretation of the OVRO lightcurves on the shortest observed timescales, e.g., in multiwavelength studies of radio flares and jet physics like the ones referenced above. It is also important for the design of future surveys to study the radio variability of AGNs (and other compact sources) with next generation radio telescopes such as the Square Kilometre Array \citep{bignalletal15} and its precursors \citep{murphyetal13}, where being able to distinguish between both forms of variability is needed to understand the underlying physics. 

In this paper, we investigate the origin of the 15\,GHz variability of the OVRO-monitored blazars on the shortest observed timescale of $\sim 4$ days. We use the term interday variability to define flux density variations occurring on a timescale of days. This is the first ever study of interday variability at 15\,GHz for such a large sample of sources. We describe the source sample briefly in Section~\ref{sample}, then characterize the 4-day variability amplitudes using the structure function in Section~\ref{characterization}. In Section~\ref{results}, we determine if ISS is responsible for the interday variability of these OVRO blazars by examining the Galactic dependence of their variability amplitudes, and discuss the implications of our results on blazar interday variability at 15\,GHz. A summary of the paper is provided in Section~\ref{summary}.

\section{Source Sample}\label{sample}

For this study, we use the original sample of 1158 sources monitored by the OVRO 40-m telescope (Richards et al. 2011), selected from the Candidate Gamma-Ray Blazar Survey (CGRaBS, Healey et al. 2008). CGRaBS sources above a declination cut of $> -20^{\circ}$ were selected for monitoring by the OVRO telescope. The original CGRaBS sample was selected such that the sources would have spectral indices, radio flux densities and X-ray flux densities similar to those of Energetic Gamma Ray Experiment Telescope (EGRET) detected sources, and would thus have a high chance of being detected in gamma-rays by \textit{Fermi}. The CGRaBS sources were also selected to be outside $\pm 10^{\circ}$ of the Galactic plane.

The OVRO telescope has been monitoring these sources at a cadence of around twice per week since 2008 to the present, subject to weather conditions and the instrument being operational. Additionally, about 20\% of the sources in the OVRO sample would be randomly selected each week to be observed only once that week, to fit into the schedule. Therefore, while the median time sampling of each source is about 4 days, the time lag between consecutive flux measurements in the OVRO lightcurves can be $\sim$8 days or more. For our analysis, we include flux density measurements up till 2018 April 10. 

\citet{richardsetal11} provide a detailed description of the observations and data reduction methodologies of the OVRO program.

\section{Characterization of variability amplitudes}\label{characterization} 

\subsection{The structure function}\label{sf} 

\begin{figure}
	\begin{center}
		\includegraphics[width=\columnwidth]{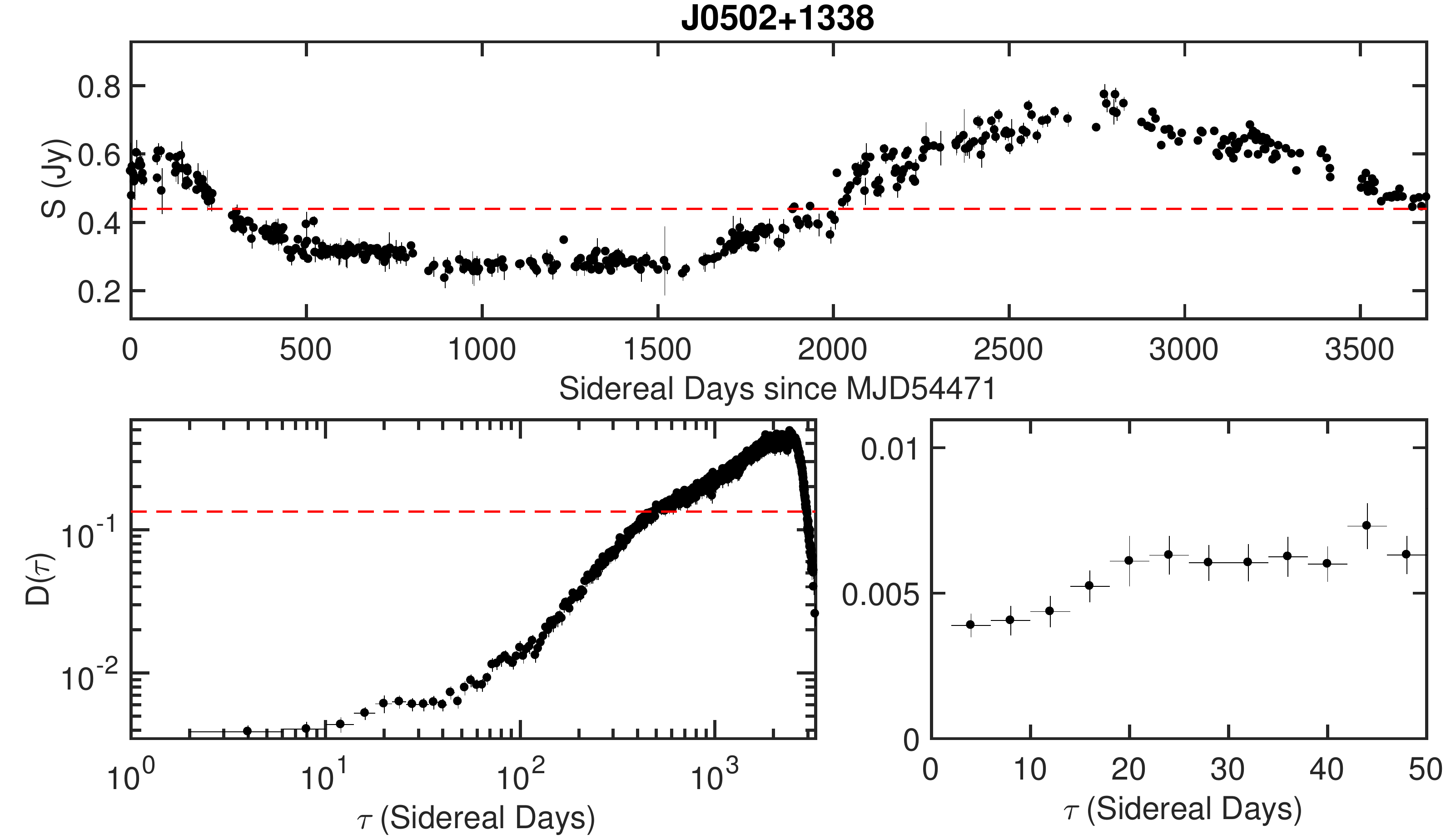}
	\end{center}
	\caption{{Top: Lightcurve for the source J0502+1338, where the horizontal dashed line denotes the mean flux density of the source. The error bars are given by Equation~\ref{erroreqrichards} \citep{richardsetal11}. Bottom: Structure function, $D(\tau)$, calculated from the lightcurve using Equation~\ref{sfeq}, shown in its entirety in the left panel, and for $\tau \leq 50 \rm d$ in the right panel. The horizontal dashed line denotes $D_{m15}$ (Equation~\ref{m2D}) derived from the intrinsic modulation indices estimated by \citet{richardsetal14}. \label{J0502}}}
\end{figure}

We use the structure function amplitude to characterize the strength of variability at different timescales, given as: 
\begin{equation}
D(\tau) = \dfrac{1}{N_{\tau}} \sum_{j,k} \left( \dfrac{S_j - S_k}{S_{15}} \right) ^{2} \label{sfeq}
\end{equation}
where $S_j$ and $S_k$ represent a pair of measured flux densities separated by a time interval  $\tau$, binned to the nearest integer multiple of 4 days. $S_{\rm 15}$ is the mean flux density calculated over the full lightcurve. $N_{\tau}$ is the number of pairs of flux densities in each time lag bin. We selected bins in integer multiples of $\tau = 4 \rm \,d$ since it is the typical smallest time lag between successive data samples in the OVRO program for the majority of the sources. We note that $N_{\tau}$ typically decreases with increasing $\tau$, with $N_{\rm 4d} \approx 2 N_{\rm 8d}$, and so on. Bins were thus selected for plotting $D(\tau)$ and for our analysis only if $N_{\tau} \geq 30$. An example of a source lightcurve and the corresponding structure function is shown in Figure~\ref{J0502} for the source J0502+1338. The error bars for $D(\tau)$ shown in the bottom panels of Figure~\ref{J0502} are estimated as the standard error in the mean, defined as the ratio of the standard deviation of the $[(S_j - S_k)/S_{15}]^{2}$ terms in that particular time lag bin to $\sqrt{N_{\tau}-1}$. This error estimate does not take into account the statistical errors due to the finite span of the OVRO observations, which would increase as $\tau$ increases relative to the total observing timespan. 

\begin{figure}
	\begin{center}
		\includegraphics[width=\columnwidth]{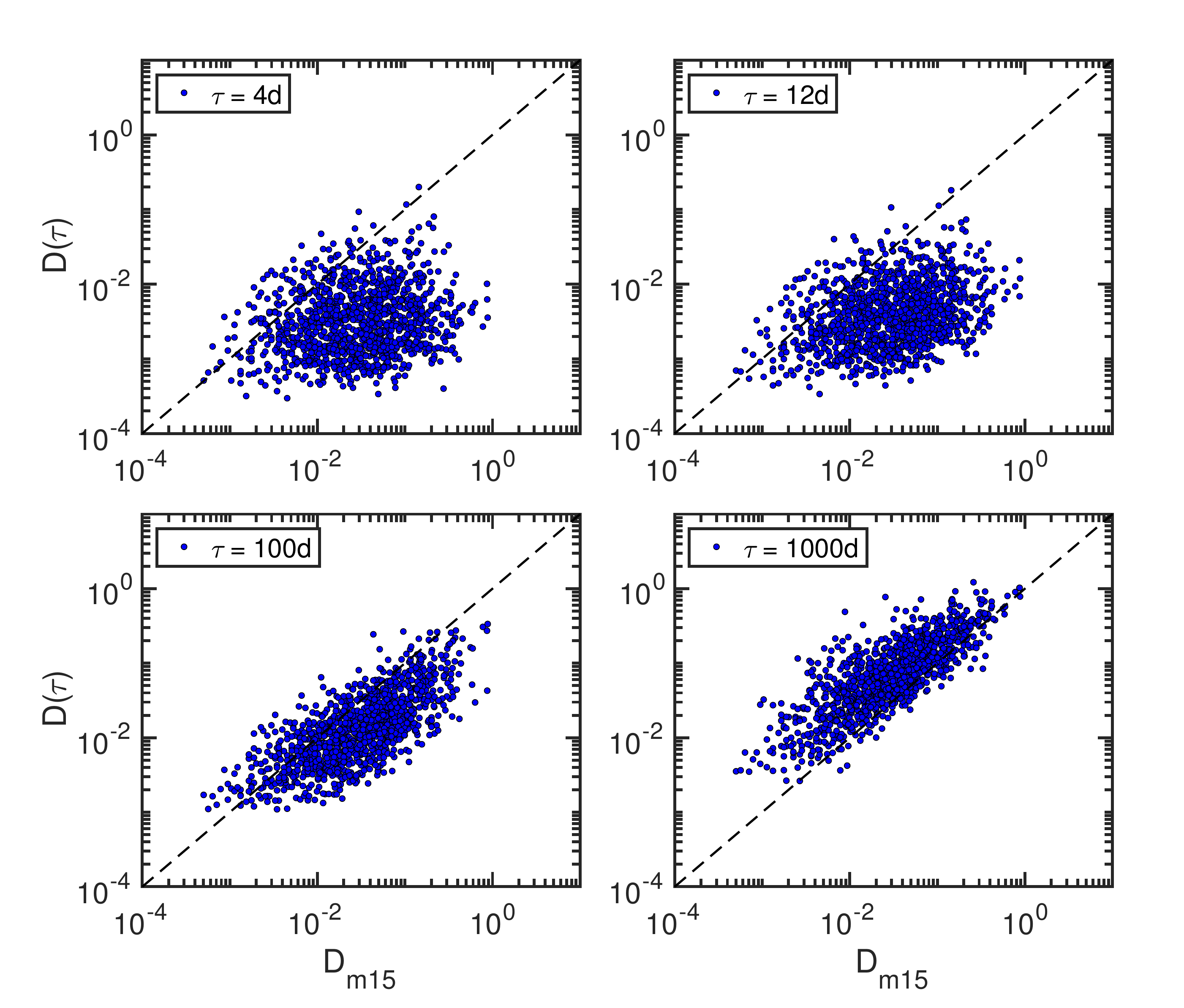}
	\end{center}
	\caption{{Structure function amplitudes, $D(\tau)$, for $\tau = 4, 12, 100, 1000$ days, plotted against $D_{m15}$ derived using Equation~\ref{m2D} from the intrinsic modulation indices, $m_{15}$, published by \citet{richardsetal14}. The dashed line shows the $x = y$ line. \label{sfvsm15}}}
\end{figure}

As a sanity check, we compare $D({\tau})$ against the intrinsic modulation index, $m_{15}$, as determined using the maximum likelihood method by \citet{richardsetal14}. Since $m_{15}$ is a measure of the standard deviation, whereas $D({\tau})$ is a measure of the variance, we convert $m_{15}$ to an equivalent structure function amplitude following:
\begin{equation}
D_{m15} = 2(m_{15})^{2} \,\,\, , \label{m2D}
\end{equation}
based on the assumption that the structure function amplitudes have saturated. Figure~\ref{sfvsm15} shows that $D({\tau})$ approaches and becomes comparable to $D_{m15}$ as $\tau$ increases to of order 100 to 1000 days. This confirms that $m_{15}$ is more representative of the variability amplitude on timescales of a hundred days or longer. We note that the $D({\tau})$ values shown here were derived from lightcurves in which outliers have been flagged (described in Section~\ref{noise} below). Also, the $m_{15}$ values derived by \citet{richardsetal14} were based only on the first 4 years of the OVRO data.

\subsection{Data flagging and error estimation}\label{noise}

\begin{figure}
	\begin{center}
		\includegraphics[width=\columnwidth]{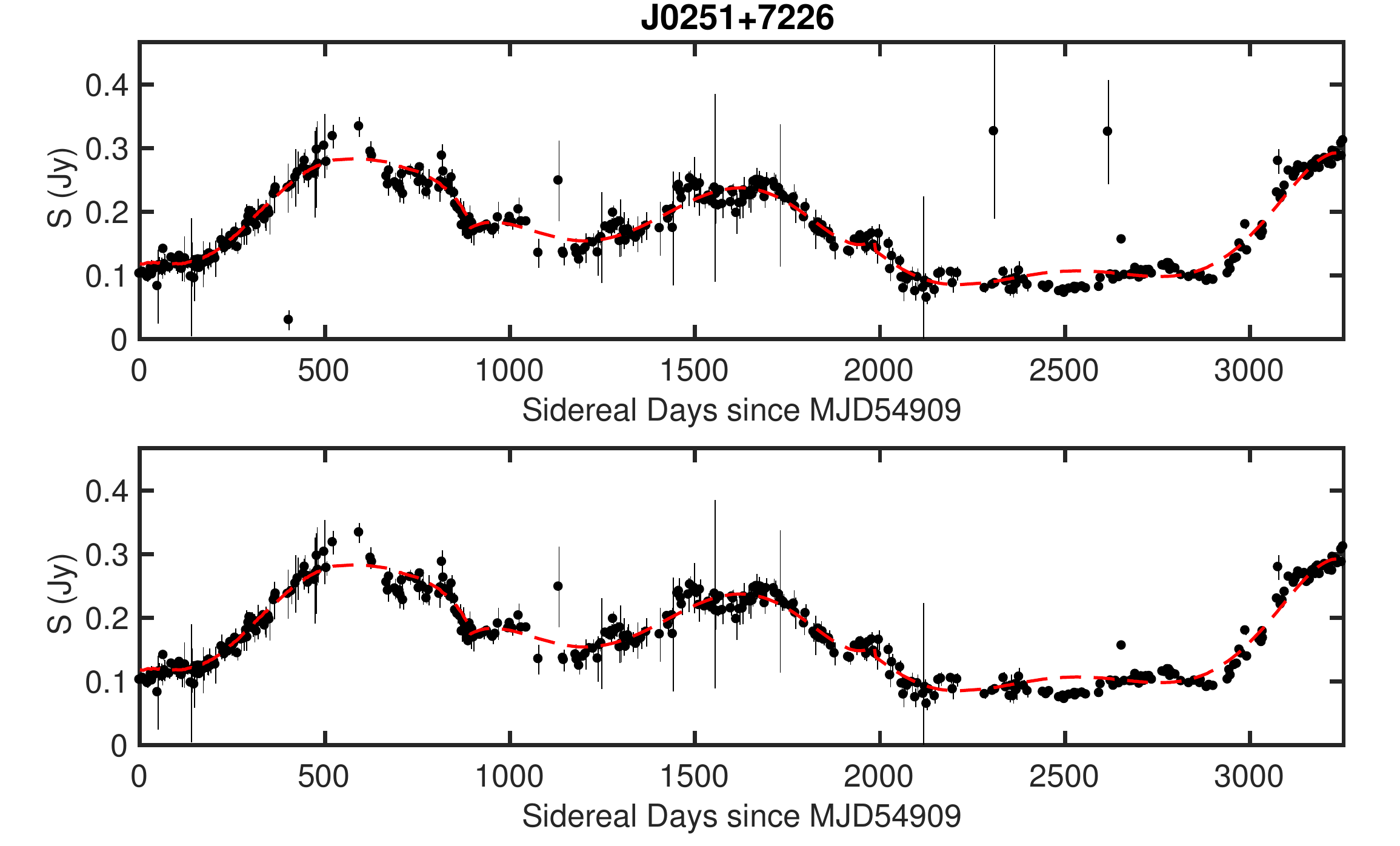}
	\end{center}
	\caption{{Example of automated flagging of the lightcurves for the source J0251$+$7226. Polynomial functions (dashed curves) are fit to 3 contiguous segments, then data points for which the fit residuals are $\geq 4$ times the rms residuals are flagged. The top panel shows the lightcurve prior to flagging while the bottom panel shows the lightcurve after 3 outlier datapoints have been flagged automatically. \label{flagexample}}}
\end{figure}

Many of the OVRO lightcurves contain outliers that skew the structure function amplitudes. To automatically flag off these outliers, we first divided each source lightcurve into 3 contiguous segments of equal time period, then fit a 6th order polynomial to each segment. This segmentation enables better fits to the lightcurves, particularly those that exhibit rapid variations with many inflections over the full 10 year period. We then remove datapoints for which the residuals are $\geq 4$ times that of the rms residuals over the corresponding segment. An example of this automatic flagging is shown in Figure~\ref{flagexample}, for the source J0251$+$7226.

Errors in flux density measurements due to instrumental and other systematic effects contribute to the measured $D(\tau)$. One can be very conservative and assume that the flux density variations on the shortest measured timescales, as characterized by $D(\rm 4d)$, provides an upper limit on such errors in the flux density measurements. However, using $D(\rm 4d)$ will overestimate the errors particularly in sources that exhibit real variability (whether ISS or intrinsic) on these short timescales.  

Since our goal is to examine if ISS is present in $D(\rm 4d)$, we use instead the uncertainty of each single flux density measurement, described in \citep{richardsetal11} and given by:
\begin{equation}
\sigma_{\rm err} = \sqrt{\sigma^2_{\rm 15} + (\epsilon \cdot {S})^2 + (\eta \cdot \psi)^2 } \label{erroreqrichards}
\end{equation}
where $\sigma_{\rm 15}$ is the scatter during each flux density measurement, and accounts for thermal noise, atmospheric fluctuations and other stochastic errors. $\epsilon$ accounts for all the flux-dependent errors, including pointing and tracking errors. $\psi$ is the switched power, and the $\eta$ term accounts for systematic effects between the different beam switching pairs in each observation, caused by rapid atmospheric variations or pointing errors. The values of $\epsilon$ and $\eta$ were determined from data of sources that show little or very slow variations, using the fitting methods described in \citet{richardsetal11}. These were checked for different observing epochs, and large changes were seen, for example, when the receiver was upgraded in May 2014. The value of $\epsilon$ depends strongly on whether the source was used as a pointing source (with values ranging from $0.006-0.017$) or if it was observed within 15$\degr$ of a pointing source (classified as an `ordinary source' with values between $0.014-0.036$), with the former showing expectedly smaller pointing induced errors. The value of $\eta$ is also seen to differ between pointing sources (values between $1.22-2.24$) and ordinary sources (values between $0.47-1.59$), showing that the switched power measurements also have a dependence on flux density, as the pointing sources are typically brighter than the ordinary sources. 

As described in \citet{richardsetal11}, in some cases it is evident that the values of $\eta$ and $\epsilon$ result in too large uncertainties for some objects, which clearly show common long-term trends with scatter about the mean smaller than expected from the error model. In order to account for this effect, a cubic spline fit was used to determine a scaling factor that is then applied to scale the uncertainty due to the flux density and switched power (see \citet{richardsetal11} for details). This was not applied to the data taken after the receiver upgrade in May 2014 so that some of the uncertainties in the data may still be overestimated.

$\sigma_{\rm err}$ in Equation~\ref{erroreqrichards} also does not include the uncertainty introduced by the flux density calibration, due to possible variability of the flux calibrator sources. This is typically assumed to be $\sim5\%$ based on the observed long-term variability of the flux calibrators, but is expected to be lower on interday timescales. We estimate the flux calibration errors on 4-day timescales to be $\sim 1\%$ of the source mean flux density; the justification for this value is described in Appendix~\ref{fluxcalibrationerrors}. 

For each source, we thus estimate the total contribution of noise, calibration and other systematic errors to the observed 4-day modulation indices as the quadratic sum of the median value of $\sigma_{\rm err}$ and the $\sim 1\%$ flux calibration errors, normalized by the mean flux density (see Equation~\ref{msigmawithfceq} in Appendix~\ref{fluxcalibrationerrors}):
\begin{equation}
m_{\sigma} = \dfrac{\sqrt{({\rm median}(\sigma_{\rm err}))^2 + (0.01S_{\rm 15})^2}}{S_{\rm 15}}.  \label{msigmaeq}
\end{equation}
The rationale behind Equation~\ref{msigmaeq} is that the total error estimate determines how much the flux densities can vary from one measurement to the next, in the absence of real astrophysical variability; $m_{\sigma}$ thus represents the estimated error contribution to the variability amplitudes on the shortest observed timescales. We use the median instead of the mean $\sigma_{\rm err}$ value, since the presence of a few large $\sigma_{\rm err}$ in a lightcurve (as can be seen in Figures~\ref{J0502} and \ref{flagexample}) skews the mean towards larger values, which in turn may overestimate the errors. As a check, when we use the mean instead of the median $\sigma_{\rm err}$ value to estimate $m_{\sigma}$, we find that the distribution of $m_{D({\rm4d})}/m_{\sigma}$ peaks at values $< 1$, where $m_{D(\rm 4d)}$ is the modulation index derived from $D(\rm 4d)$ using Equation~\ref{m2D}; this suggests that using the mean of $\sigma_{\rm err}$ overestimates $m_{\sigma}$ for each source.

\begin{figure}
	\begin{center}
		\includegraphics[width=\columnwidth]{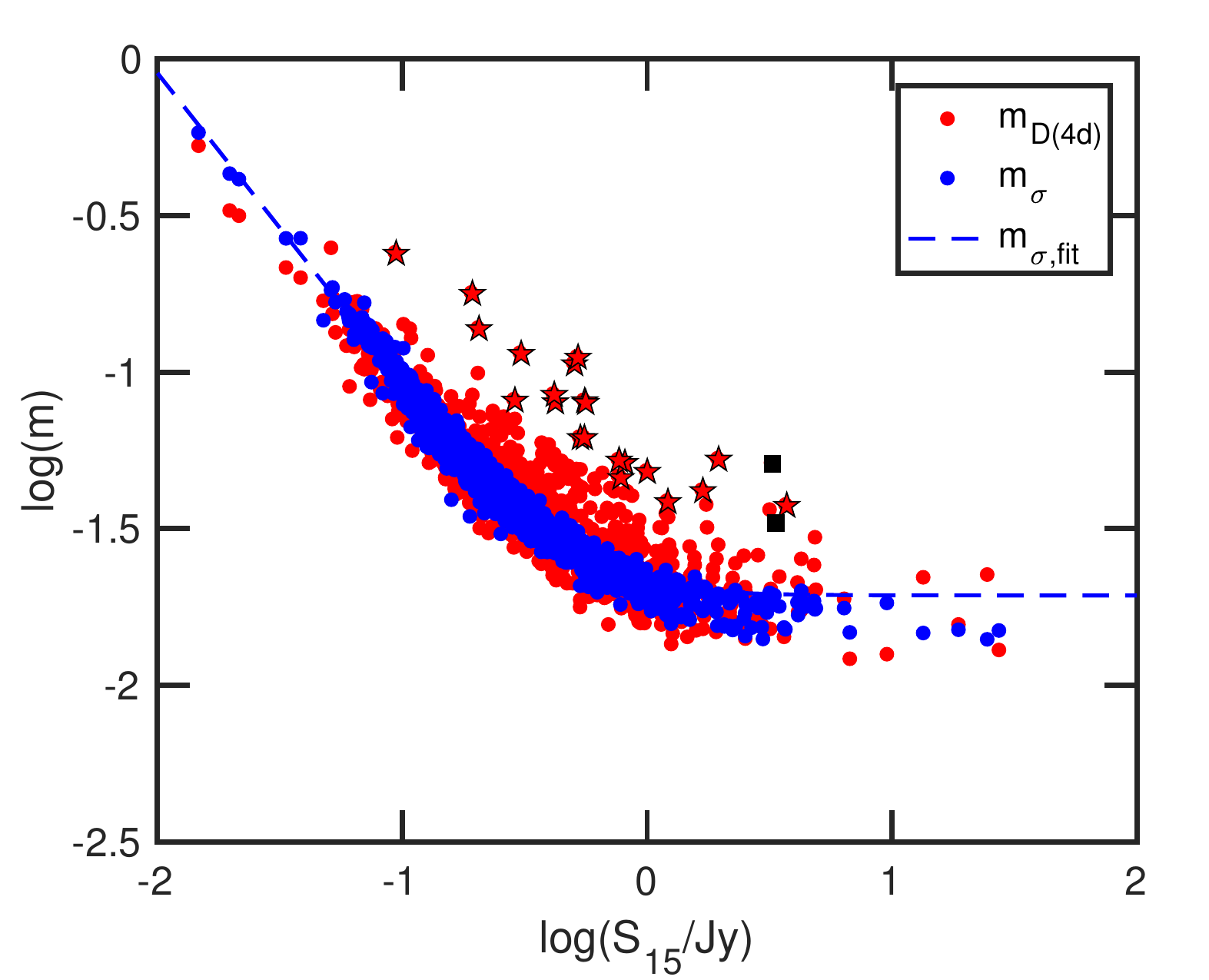}
	\end{center}
	\caption{{Modulation indices derived from the 4-day structure function amplitude, $m_{D(\rm 4d)}$, and the total contribution of instrumental, calibration and other systematic errors to the observed 4-day modulation indices of each source, $m_{\sigma}$, plotted against the mean 15\,GHz flux density. The red star symbols denote sources for which $m_{D(\rm 4d)} \geq 2m_{\sigma}$. The dashed line denotes the best fit of Equation~\ref{erroreq} to $m_{\sigma}$. The black squares show $m_{D(\rm 4d)}$ for two blazars observed through the Galatic plane that were not included in our sample but were also monitored by OVRO since 2008, i.e., 3EGJ\,2016$+$3657 and 3EGJ\,2027$+$3429 (see Section~\ref{galacticplaneblazars}). \label{mvsS15}}}
\end{figure}

A diagnostic plot of $m_{D(\rm 4d)}$ (in red) vs. 15\,GHz mean flux density is shown in Figure~\ref{mvsS15}. Overlayed are plots of $m_{\sigma}$ (in blue) for each source. The dashed line shows the following fit to $m_{\sigma}$:
\begin{equation}
 m_{\sigma,\rm fit} = \sqrt{p^2 + ({s}/{S_{\rm 15}})^2} \label{erroreq}
\end{equation}
where $s$ collates all the flux independent errors, i.e., $\sigma_{15}$ and $\eta \cdot \psi$ in Equation~\ref{erroreqrichards}, while $p$ collates all the flux dependent errors. We obtained best fit values of $p = 0.0194$ and $s = 0.009$\,Jy for $m_{\sigma}$.

\begin{figure}
	\begin{center}
		\includegraphics[width=\columnwidth]{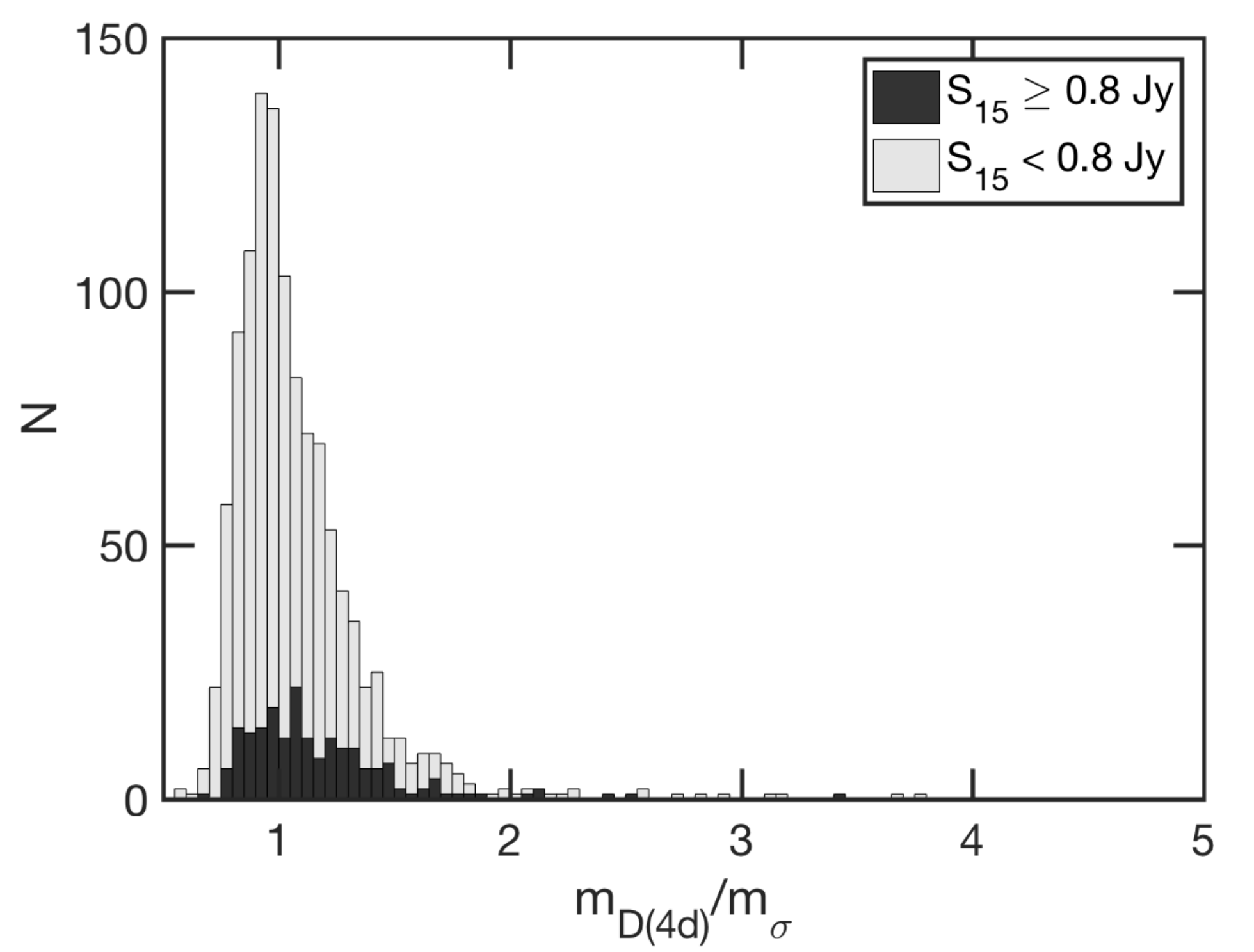}
	\end{center}
	\caption{{Histogram showing the distribution of the ratio of the 4-day variability amplitudes to the flux normalized measurement uncertainties of each source, $m_{D({\rm 4d})}/m_{\sigma}$. The histograms peak at a value of $\sim 1$ for both the $S_{\rm 15} \geq 0.8$\,Jy and $S_{\rm 15} < 0.8$\,Jy sources, when the 1\% flux calibration errors are included. The errors in the $S_{\rm 15} \geq 0.8$\,Jy sources are likely underestimated when excluding the flux calibration errors (Appendix~\ref{fluxcalibrationerrors}).  \label{diffmnoise}}}
\end{figure}

From Figure~\ref{mvsS15}, we see that $m_{D({\rm 4d})}$ is generally comparable to $m_{\sigma}$ for the large majority of sources, displaying a similar flux density dependence. This is to be expected if $m_{D({\rm 4d})}$ is dominated by noise and systematic uncertainties as characterised by Equation~\ref{erroreq} for the majority of sources. This is also demonstrated in Figure~\ref{diffmnoise} where the distribution of $m_{D({\rm4d})}/m_{\sigma}$ peaks at a value of $\sim$1, for both the $S_{15} \geq 0.8$\,Jy and $S_{15} < 0.8$\,Jy sources. As shown in Figure~\ref{mvsS15nofc} and discussed in Appendix~\ref{fluxcalibrationerrors}, not including the estimated 1\% flux calibration errors results in an underestimation of $m_{\sigma}$ for the $S_{15} \geq 0.8$\,Jy. The tail towards larger values of $ m_{D({\rm4d})}/m_{\sigma}$ ($> 1.5$) suggests the presence of real astrophysical variability in a fraction of the OVRO sources at these 4\,day timescales; 21 of the 1158 sources (1.8\%) show 4-day variability amplitudes $\geq 2$ times that of $m_{\sigma}$. We discuss the origin of this variability in the next section.

\section{Results and discussion}\label{results}

\subsection{Galactic dependence of variability amplitudes}\label{galdepend}

\begin{figure}
	\begin{center}
		\includegraphics[width=\columnwidth]{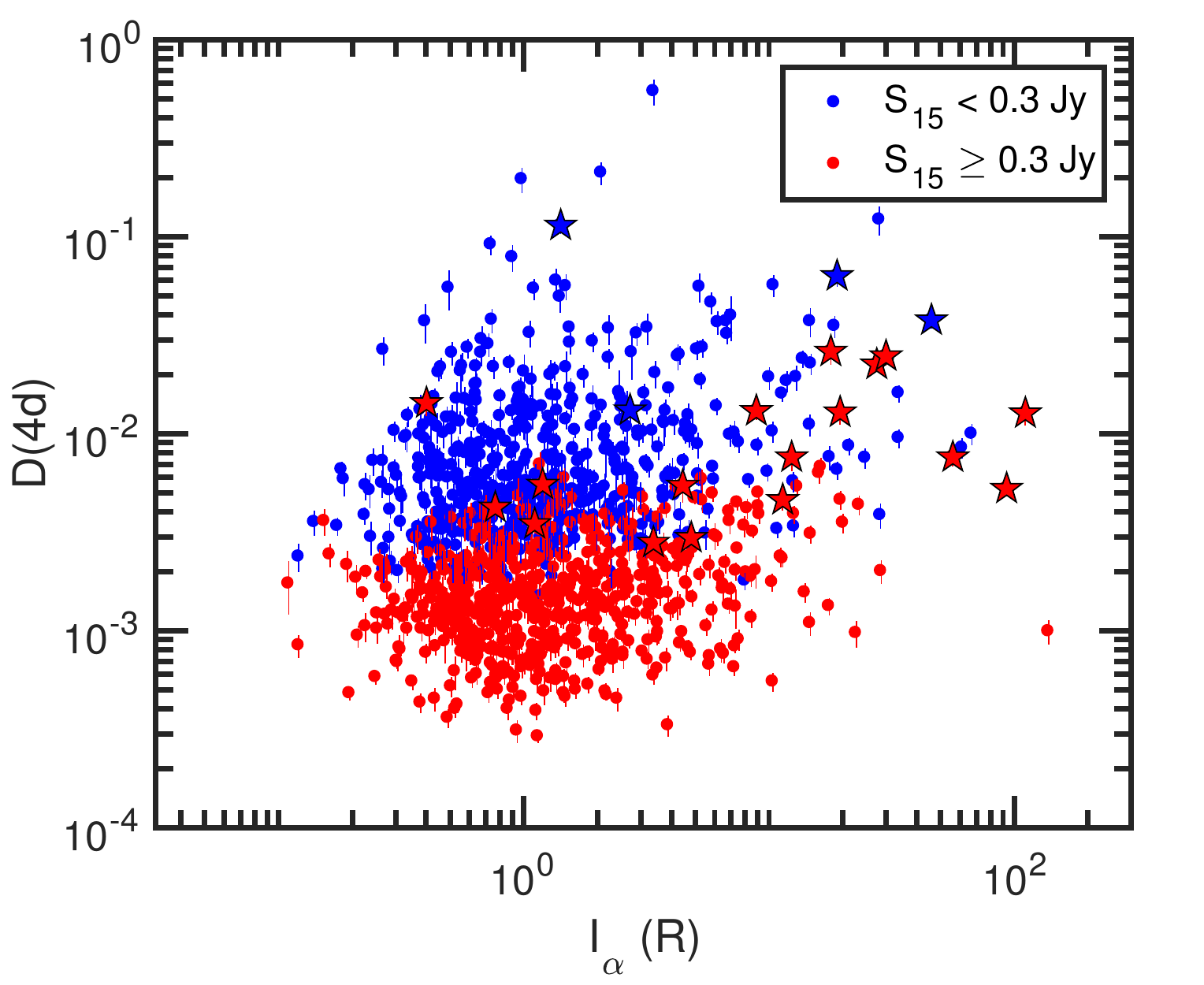}
		\includegraphics[width=\columnwidth]{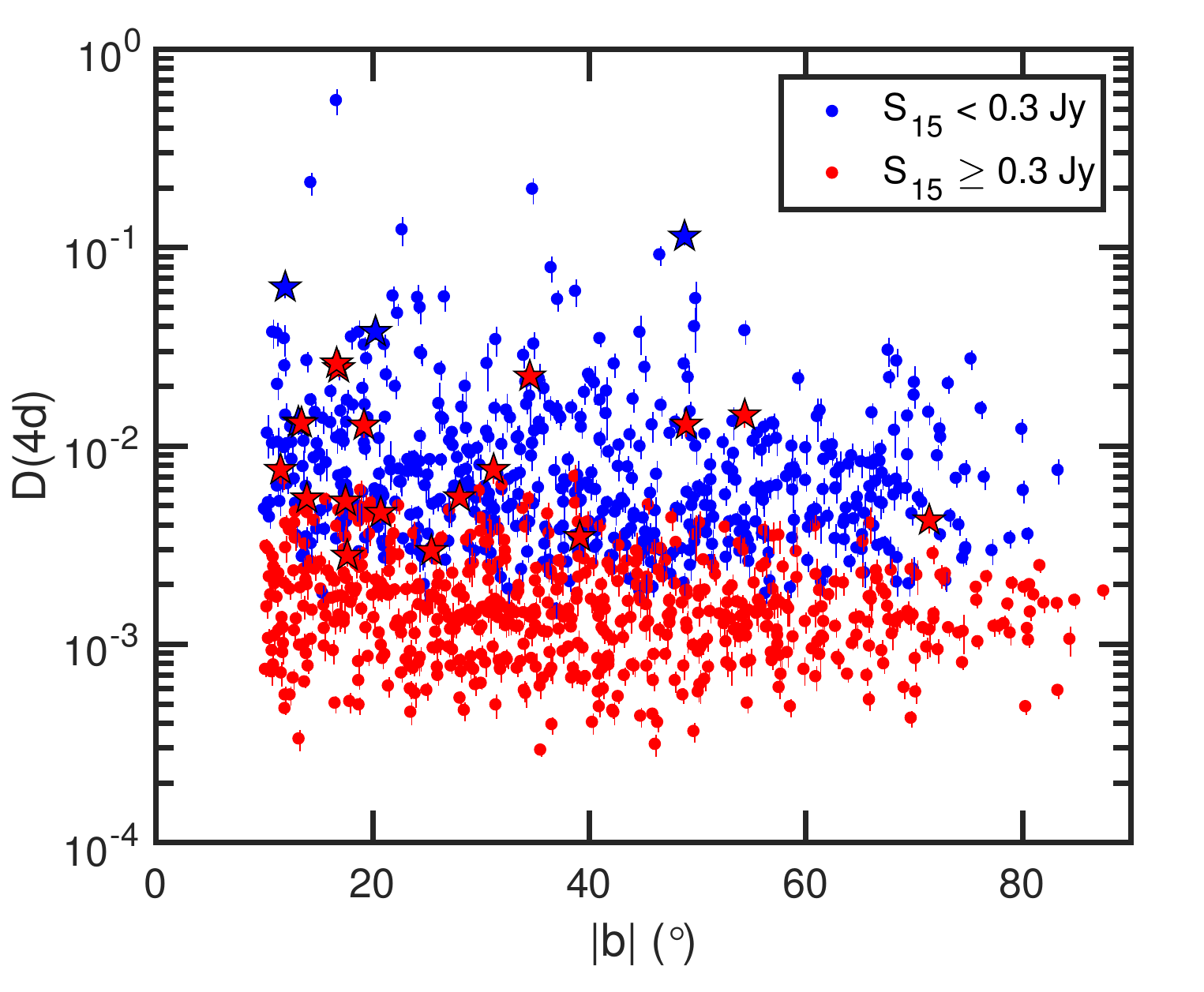}
	\end{center}
	\caption{{Structure function amplitude at 4 day timescales, $D(\rm 4d)$, vs line-of-sight H$\alpha$ intensity (top) and Galactic latitude (bottom). The sources are separated into two roughly equal samples of high flux density ($S_{\rm 15} \geq 0.3$\,Jy, red) and low flux density ($S_{\rm 15} < 0.3$\,Jy, blue) sources. The star symbols denote the most significant variables in our sample (discussed in Section~\ref{variables}). The fact that the relationship between $D(\rm 4d)$ and $I_{\alpha}$ is evident for both the weak and strong source samples confirms that this relationship is not due the presence of noise in the $D(\rm 4d)$ of the weaker sources. \label{sfhalphagalcorr}}}
\end{figure}

For our full sample of 1158 sources, we now examine if their variability amplitudes on timescales of days and weeks show a Galactic dependence, which would provide strong evidence for the presence of ISS. The top panel of Figure~\ref{sfhalphagalcorr} shows $D(\rm 4d)$ plotted against the line-of-sight H$\alpha$ intensities ($I_{\alpha}$) obtained from the Wisconsin H-Alpha Mapper (WHAM) Survey \citep{haffneretal03}. Since the H$\alpha$ intensities are a measure of the integral of the squared electron densities along the line of sight, they provide a proxy for the line-of-sight interstellar scattering strength. Indeed, the intra and interday variability amplitudes of blazars at 2\,GHz \citep{rickettetal06}, 5\,GHz \citep{lovelletal08} and 8\,GHz \citep{koayetal12} show significant correlations with line-of-sight Galactic H$\alpha$ intensities, demonstrating that their flux density variations are dominated by ISS.

For both the weak and strong source samples, there is a clear excess of sources with larger amplitude variability for sight-lines where $I_{\alpha} \geq 10$\,rayleighs (R). Spearman correlation tests show a statistically significant relationship between $D(\rm 4d)$ and the line-of-sight H$\alpha$ intensities ($p$-value of $2.67 \times 10^{-4}$), as shown in Table~\ref{spearmantable}. We have chosen a significance level of $\alpha = 0.05$. This H$\alpha$ dependence of the 15\,GHz variability amplitudes demonstrates the presence of ISS in the OVRO lightcurves, at least in sources observed through heavily scattered lines of sight. In fact, this correlation between $D(\tau)$ and $I_{\alpha}$ remains statistically significant up to a timescale of $\tau \sim 80 \rm d$ (Table~\ref{spearmantable}). However, on timescales of 100 days and above, this correlation is no longer significant as intrinsic variations likely begin to dominate. 

\begin{table*}
	\centering
	\caption{Spearman rank correlation coeffficients, $r_s$, and corresponding $p$-values between pairs of parameters and for various source samples.
		\label{spearmantable}}
	\begin{tabular}{l c c c c c c}
		\hline
		\hline
		Parameter 1 & Parameter 2 & Source sample & No. of sources & $r_s$ & $p$-value & Significant? \\
		& & &  & &  & ($\alpha = 0.05$) \\
		\hline
		$D(\rm 4d)$   &	$I_{\alpha}$	& 	all  & 1158 & 	0.107		& $2.67 \times 10^{-4}$	& Y\\
		$D(\rm 8d)$	 &  $I_{\alpha}$	& all  & 1158 &	  0.111	& 	$1.47\times 10^{-4}$	& Y\\
		$D(\rm 12d)$	 &  $I_{\alpha}$	& all  & 1158 &	 0.121	& 	$3.54 \times 10^{-5}$	& Y\\
		$D(\rm 16d)$	 &  $I_{\alpha}$	& all  & 1158 &	0.118	& 	$5.97 \times 10^{-5}$	& Y\\
		$D(\rm 20d)$	 &  $I_{\alpha}$	&  all  & 1158 &	0.112	& 	$1.35 \times 10^{-4}$	& Y\\
		$D(\rm 40d)$	 &  $I_{\alpha}$	&  all  & 1158 &	0.073	& 	$1.26 \times 10^{-2}$	& Y\\
		$D(\rm 60d)$	 &  $I_{\alpha}$	&  all  & 1158 &	0.068	& 	$2.03 \times 10^{-2}$	& Y\\
		$D(\rm 80d)$	 &  $I_{\alpha}$	&  all  & 1158 &	0.063	& 	$3.17 \times 10^{-2}$	& Y\\
		$D(\rm 100d)$	 &  $I_{\alpha}$	&  all  & 1158 &	$-$0.055	& 	$5.95 \times 10^{-2}$	& N\\
		$D(\rm 1000d)$	 &  $I_{\alpha}$	&  all  & 1158 &	$-$0.047	& 	$1.12\times 10^{-1}$	& N\\
	       \hline
		$D(\rm 4d)$   &	$I_{\alpha}$	& 	$I_{\alpha} < 10$ & 1104 & 	0.059		& $4.87 \times 10^{-2}$	& Y\\
		$D(\rm 8d)$	 &  $I_{\alpha}$	& $I_{\alpha} < 10$ & 1104 &	  0.065	& 	$3.18\times 10^{-2}$	& Y\\
		$D(\rm 12d)$	 &  $I_{\alpha}$	&  $I_{\alpha} < 10$ & 1104 &	 0.082	& 	$6.30 \times 10^{-3}$	& Y\\
		$D(\rm 16d)$	 &  $I_{\alpha}$	& $I_{\alpha} < 10$ & 1104 &	 0.078	& 	$9.20 \times 10^{-3}$	& Y\\
		$D(\rm 20d)$	 &  $I_{\alpha}$	&  $I_{\alpha} < 10$ & 1104 &	0.077	& 	$1.04 \times 10^{-2}$	& Y\\
		$D(\rm 40d)$	 &  $I_{\alpha}$	&  $I_{\alpha} < 10$ & 1104 &	0.041	& 	$1.74 \times 10^{-1}$	& N\\
		$D(\rm 60d)$	 &  $I_{\alpha}$	&  $I_{\alpha} < 10$ & 1104 &	0.036	& 	$2.39 \times 10^{-1}$	& N\\
		$D(\rm 80d)$	 &  $I_{\alpha}$	&  $I_{\alpha} < 10$ & 1104 &	0.032	& 	$2.85 \times 10^{-1}$	& N\\
		$D(\rm 100d)$	 &  $I_{\alpha}$	&  $I_{\alpha} < 10$ & 1104 &	$-$0.029	& 	$3.44 \times 10^{-1}$	& N\\
		$D(\rm 1000d)$	 &  $I_{\alpha}$	&  $I_{\alpha} < 10$ & 1104 &	$-$0.018	& 	$5.42\times 10^{-1}$ & N\\		
	\hline
$D(\rm 4d)$   &	$\lvert b \rvert ^{\circ}$	& all	 & 1158 & 	$-$0.072		& $1.40 \times 10^{-2}$	& Y\\
$D(\rm 8d)$	 &  $\lvert b \rvert ^{\circ}$	&  all & 1158 &	  $-$0.075		& 	$1.04 \times 10^{-2}$	& Y\\
$D(\rm 12d)$	 &  $\lvert b \rvert ^{\circ}$	&  all  & 1158 &	 $-$0.090	& 	$2.30 \times 10^{-3}$	& Y\\
$D(\rm 16d)$	 &  $\lvert b \rvert ^{\circ}$	&  all & 1158 &	$-$0.086	& 	$3.50 \times 10^{-3}$	& Y\\
$D(\rm 20d)$	 &  $\lvert b \rvert ^{\circ}$	&  all  & 1158 &	$-$0.084	& 	$4.40\times 10^{-3}$	& Y\\
$D(\rm 40d)$	 &  $\lvert b \rvert ^{\circ}$	&  all  & 1158 &	$-$0.045	& 	$1.25 \times 10^{-1}$	& N\\
$D(\rm 60d)$	 &  $\lvert b \rvert ^{\circ}$	&  all  & 1158 &	$-$0.038	& 	$1.94 \times 10^{-1}$	& N\\
$D(\rm 80d)$	 &  $\lvert b \rvert ^{\circ}$	&  all  & 1158 &	$-$0.033	& 	$2.61 \times 10^{-1}$	& N\\
$D(\rm 100d)$	 &  $\lvert b \rvert ^{\circ}$	&  all  & 1158 &	$-$0.026	& 	$3.75 \times 10^{-1}$	& N\\
$D(\rm 1000d)$	 &  $\lvert b \rvert ^{\circ}$	&  all  & 1158 &	$-$0.217	& 	$4.75\times 10^{-1}$	& N\\
		\hline
		\hline
	\end{tabular}
\end{table*}

The Spearman correlation tests may be biased by the extreme $I_{\alpha} \geq 10\, \rm R$ sources. We therefore repeat the same tests using only sources with line-of-sight $I_{\alpha} < 10\, \rm R$. We find that the correlation between $D(\tau)$ and $I_{\alpha}$ remains significant, up to a timescale of $\sim$20 days. This suggests that at 15\,GHz, while the variability of sources seen through heavily scattered sight-lines ($I_{\alpha} \geq 10\, \rm R$) may be dominated by ISS up to timescales of 80 days, ISS is significant up to only $\sim$20 day timescales for more typical sightlines through the Galaxy where $I_{\alpha} < 10\, \rm R$.

\begin{figure}
	\begin{center}
		\includegraphics[width=\columnwidth]{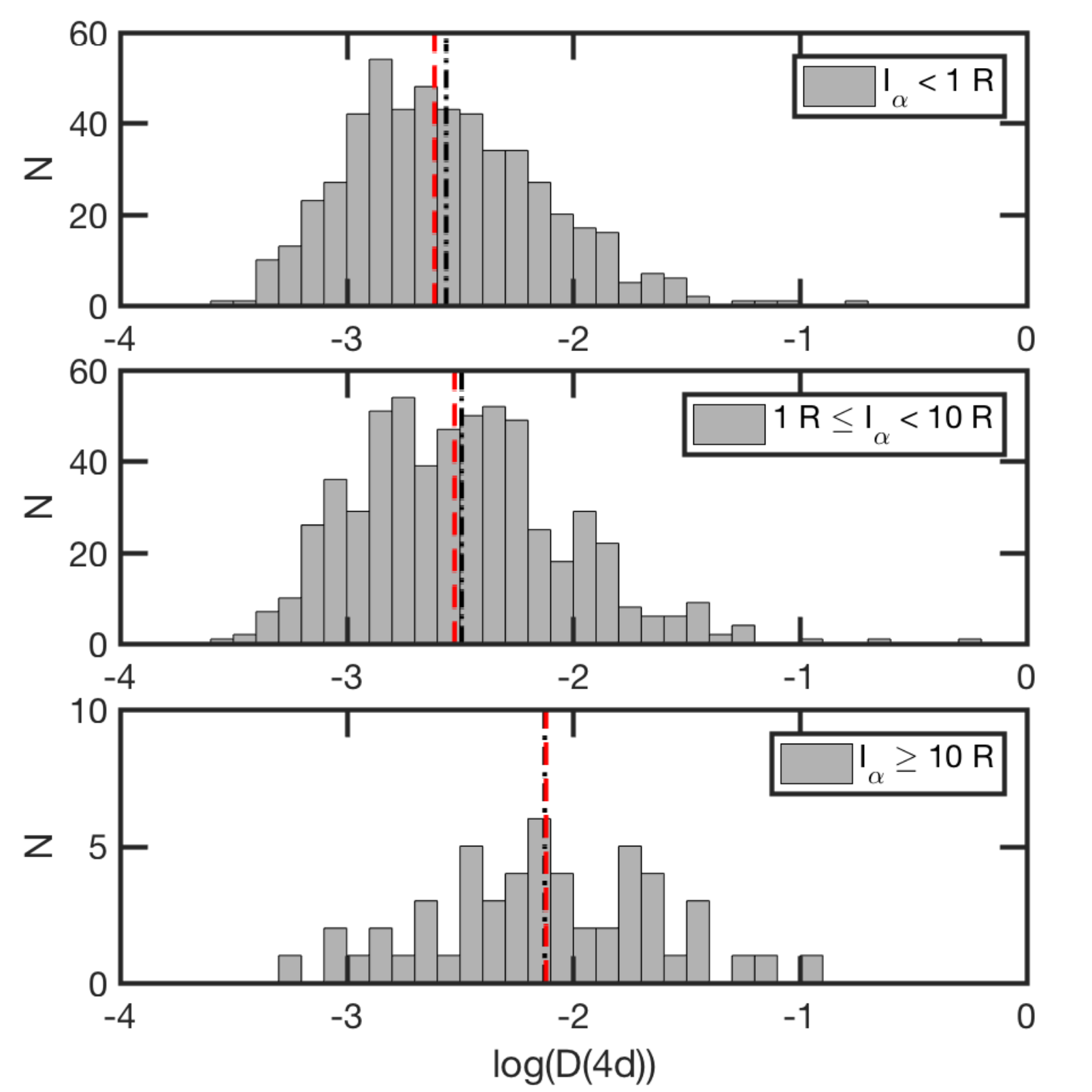}
	\end{center}
	\caption{{Histograms showing the distributions of $D(\rm 4d)$ for sources with low ($I_{\alpha} < 1\, \rm R$, top), moderate ($ 1\, {\rm R} \leq  I_{\alpha} < 10\, \rm R$, middle), and high ($I_{\alpha} \geq 10\, \rm R$, bottom) line-of-sight H$\alpha$ intensities. The dashed (red) and dash-dotted (black) vertical lines show the median and mean values of $D(\rm 4d)$ respectively. The $I_{\alpha} \geq 10\, \rm R$ sample of sources contain a significantly higher fraction of interday variables. \label{sfhalphadist}}}
\end{figure}

As further confirmation, we examine in Figure~\ref{sfhalphadist} the distribution of $D(\rm 4d)$ for sources with low ($I_{\alpha} < 1\, \rm R$, top), moderate ($ 1\, {\rm R} \leq  I_{\alpha} < 10\, \rm R$, middle), and high ($I_{\alpha} \geq 10\, \rm R$, bottom) line-of-sight H$\alpha$ intensities. The Kolmogorov-Smirnov (K-S) test confirms that the distribution of $D(\rm 4d)$ for sources with high $I_{\alpha}$ is significantly different from that of the combined sample of sources with low and moderate $I_{\alpha}$, at a $p-$value of $6.45 \times 10^{-6}$. The mean value of $D(\rm 4d)$ for sources with $I_{\alpha} \geq 10\, \rm R$ is 0.0143, a factor of $\sim 2$ higher than the value of 0.0061 for that of sources with $I_{\alpha} < 10\, \rm R$.

Although we see no obvious correspondence between $D(\rm 4d)$ and the Galactic latitudes by eye (Figure~\ref{sfhalphagalcorr}, bottom), the Spearman correlation test reveals a statistically significant anti-correlation between $D(\tau)$ and $\lvert b \rvert$ on timescales of 4 to 20 days (Table~\ref{spearmantable}). The correlation coefficients are weaker compared to that between $D(\rm 4d)$ and $I_{\alpha}$. The H$\alpha$ intensities are therefore a better indicator of line-of-sight scattering strength compared to the Galactic latitudes, due to the complex structure of the ionized gas in the Galaxy. This is in spite of the $1^{\circ}$ angular resolution of the WHAM Survey data.

\subsection{ISS of the most significant interday variables }\label{variables}

Since $D(\rm 4d)$ still comprises significant amounts of instrumental and systematic errors in a large fraction of sources (i.e., the peak of $m_{D(\rm 4d)}/m_{\sigma}$ is close to unity), we now examine only the most significant variables at 4-day timescales to determine the origin of their variability. We consider sources satisfying the criteria that $m_{D(\rm 4d)} \geq 2m_{\sigma}$ to be significantly variable, based on the tail end of the $m_{D(\rm 4d)}/ m_{\sigma}$ distribution in Figure~\ref{diffmnoise}. We initially find 21 significant interday variables that meet this criteria. After careful inspection (described in Appendix~\ref{j0259}), we found that the lightcurve of J0259$-$0018, the weakest ($\sim0.1$\,Jy) and most variable ($m_{D(\rm 4d)} \sim 24\%$) of these 21 sources, was severely affected by an error in source coordinates in the OVRO and CGRaBS catalogues. We therefore remove it from our sample of significant interday variables and refer to the remaining 20 sources as `interday variables' for the rest of this paper. The full list of these interday variables is shown in Table~\ref{2sigmavariableslist}, together with their variability amplitudes. Their lightcurves are presented in Appendix~\ref{appendixvariables}. 

\begin{table*}
	\caption{List of the interday variables and their mean flux densities, variability amplitudes, and line-of-sight H$\alpha$ intensities. 
		\label{2sigmavariableslist}}
	\begin{tabular}{l c c c c c c c c}
		\hline
		\hline
		Name & $S_{15}$ & $D_{\rm 4d}$ & $m_{D({\rm 4d})}$ & $m_{\sigma}$ & $m_{D({\rm 4d})}/m_{\sigma}$ & $l$ & $b$ &$I_{\alpha}$ \\
		(J2000) & (Jy) &  &  &  & & ($^\circ$) & ($^\circ$) & (R)  \\
		\hline
J0050$-$0929 & 0.781 & 4.22e$-$03 & 0.046 & 0.020 & 2.29 & 122.35 & $-$71.39 &   0.77 \\ 
J0128$+$4901 & 0.556 & 1.30e$-$02 & 0.081 & 0.027 & 3.02 &129.10 & $-$13.41 &   8.88 \\  
J0238$+$1636 & 1.689 & 3.48e$-$03 & 0.042 & 0.014 & 2.95 &156.77 & $-$39.11 &   1.11 \\ 
J0336$-$1302 & 0.422 & 1.28e$-$02 & 0.080 & 0.030 & 2.71 &201.14 & $-$48.94 &  19.49 \\ 
J0401$+$0413 & 0.505 & 2.25e$-$02 & 0.106 & 0.027 & 3.92 &186.03 & $-$34.49 &  27.49 \\ 
J0407$+$0742 & 0.534 & 7.59e$-$03 & 0.062 & 0.025 & 2.47 &183.87 & $-$31.16 &  12.37 \\ 
J0449$+$1121 & 1.001 & 4.61e$-$03 & 0.048 & 0.021 & 2.28 &187.43 & $-$20.74 &  11.37 \\ 
J0527$+$0331 & 0.522 & 2.47e$-$02 & 0.111 & 0.028 & 3.99 &199.79 & $-$16.85 &  29.99 \\ 
J0529$-$0519 & 0.206 & 3.77e$-$02 & 0.137 & 0.053 &2.61  &207.68 & $-$20.25 &  45.94 \\ 
J0541$-$0541 & 0.811 & 5.26e$-$03 & 0.051 & 0.022 & 2.32 &208.75 & $-$17.48 &  93.06 \\ 
J0542$-$0913 & 0.561 & 1.27e$-$02 & 0.080 & 0.025 & 3.15 &213.12 & $-$19.18 & 110.97 \\ 
J0552$+$0313 & 0.554 & 7.59e$-$03 & 0.062 & 0.026 & 2.41&203.23 & $-$11.48 &  56.16 \\ 
J0610$-$1847 & 0.306 & 2.61e$-$02 & 0.114 & 0.035 & 3.28 &224.10 & $-$16.66 &  17.87 \\ 
J0616$-$1041 & 0.193 & 6.32e$-$02 & 0.178 & 0.055 & 3.21 &217.39 & $-$11.91 &  18.94 \\ 
J0721$+$7120 & 1.959 & 5.52e$-$03 & 0.053 & 0.012 & 4.51 &143.98 & 28.02 &   1.19 \\ 
J0725$+$1425 & 0.768 & 5.45e$-$03 & 0.052 & 0.021 & 2.49 &203.64 & 13.91 &   4.45 \\ 
J0824$-$1527 & 0.288 & 1.32e$-$02 & 0.081 & 0.036 & 2.28 &237.08 & 13.12 &   2.71 \\ 
J1135$-$0428 & 0.417 & 1.43e$-$02 & 0.085 & 0.030 & 2.85 &269.31 & 54.34 &   0.40 \\ 
J1642$-$0621 & 1.216 & 2.96e$-$03 & 0.038 & 0.015 & 2.52 &11.48 & 25.41 &   4.82 \\ 
J1751$+$0939 & 3.717 & 2.80e$-$03 & 0.037 & 0.011 & 3.37 &34.92 & 17.65 &   3.38 \\  
		\hline
	\end{tabular}
	\begin{flushleft}
		\textbf{Column notes:} (2) 15\,GHz mean flux density; (3) 4-day structure function amplitude; (4) modulation index derived from $D({\rm 4d})$ using Equation~\ref{m2D}; (5) uncertainties in flux measurements, representing the contribution of instrumental and non-astrophysical effects to the measured variability amplitudes; (6) significance of 4-day variability amplitude, as defined by the ratio of the 4-day modulation index to the uncertainties in flux measurements; (7-8) Galactic coordinates; (9) line-of-sight H$\alpha$ intensity \citep{haffneretal03}. \\
	\end{flushleft}
\end{table*}

The flux density variations of these interday variables are clearly dominated by ISS, as evidenced by the larger fraction of variables detected among sources with larger $I_{\alpha}$ values. 11 (21\%) of the 53 sources with line-of-sight $I_{\alpha} \geq 10$\,R are classified as interday variables, while only $\sim 0.8\%$ (9/1104) of sources with $I_{\alpha} < 10$\,R are classified as such. 

ISS arises due to the scattering of radio waves by density inhomogeneities of the free electrons in the ionized interstellar medium. This scattering process is often well-described as being confined to a single thin scattering screen located between the source and the observer; this screen changes the phases of an incoming plane wave \citep{narayan92}. Compact AGN are known to exhibit ISS in two different regimes \citep{narayan92, rickett90}, weak ISS and strong refractive ISS \citep{blandfordetal86,rickett86,colesetal87}. In the weak ISS regime, phase changes in the wavefront due to diffractive scattering are less than a radian, so that the scintillation pattern is dominated by the Fresnel scale, $r_{\rm F} = \sqrt{cD_{\rm L}/(2{\pi}\nu)}$, where $c$ is the speed of light, $D_{\rm L}$ is the distance from the observer to the scattering screen, and $\nu$ is the observing frequency. In other words, weak ISS is observed at frequencies and sight-lines where the diffractive length-scale, $r_{\rm diff}$, is much larger than $r_{\rm F}$. On the other hand, the focussing and defocussing of coherent patches of waves due to the large-scale density fluctuations on the scattering screen lead to strong refractive ISS, observed when $r_{\rm diff} \ll r_{\rm F}$. ISS amplitudes are typically strongest at the transition frequency, $\nu_{0}$, between weak and strong ISS \citep{narayan92}. The modulation index of flux density variations scale as $(\nu_{0} / \nu)^{17/12}$ for weak ISS (observed at $\nu \gg \nu_{0}$) and scale as $(\nu / \nu_{0})^{17/30}$ for strong refractive ISS \citep{walker98} observed at $\nu \ll \nu_{0}$. These assume a Kolmogorov power spectrum of turbulence, as typically observed for the interstellar medium of the Milky Way \citep{goldreichsridhar95,armstrongetal95}.

Analytical solutions for the spatial coherence function of flux densities, $\Gamma_{4}(r;\nu)$, measured at two locations on the Earth separated by a distance $r$, are provided by e.g., \citet{colesetal87} and \citet{narayan92} for both the weak and strong ISS regimes. However, there are no analytical solutions for the intermediate scintillation regimes relevant for our study, where $\nu \approx \nu_{0}$. We therefore use the \citet{goodmannarayan06} fitting function derived from numerical simulations for calculations of $\Gamma_{4}(r;\nu)$ for our ensuing discussions, which is applicable at observing frequencies close to the transition frequency. By assuming that the density fluctuations on the scattering screen (and hence phase fluctuations imprinted on the scattered wave) are frozen and do not change on the timescales of interest, we can estimate theoretical values of $D({\rm 4d}) = 2[\Gamma_{4}(0;\nu) - \Gamma_{4}(r = v_{s} \cdot {\rm 4d} ; \nu)]$, where $v_{s}$ is the relative transverse velocity between the scattering screen and the Earth. 

At mid to high Galactic latitudes where $\nu_{0}$ is typically 5 to 8\,GHz \citep{walker98,walker01}, a source must contain a compact component of angular size $\theta \sim 50$ to $200\,\mu$as to scintillate at amplitudes of 2 to 10\% at 5\,GHz, as observed in the MASIV Survey \citep{lovelletal08}. At similar lines-of-sight with comparable scattering strengths, we expect sources with the same range of angular sizes to exhibit weak ISS with modulation indices of 0.5 to 2\% at 15\,GHz, as inferred from the \citet{goodmannarayan06} fitting function described above. We have assumed fiducial scattering screen distances of $D_{\rm L} = 500\,$pc and transverse velocities of $v_{\rm s} = 30{\rm \,km\,s^{-1}}$.

With the exception of 1 source exhibiting $18\%$ flux density variations, the other interday variables that we detect at 15\,GHz exhibit 3 to 13\% flux density variations on 4-day timescales, comparable to the typical ISS amplitudes observed at 5\,GHz. This higher than expected 15\,GHz ISS amplitudes can be explained if the interday variables are more compact (with $\theta \sim 5$ to $40\,\mu$as) than the typical source observed in the MASIV Survey ($\theta \sim 50$ to $200\,\mu$as) at 5\,GHz. This is in excess of the source size-frequency relation of $\theta \propto \nu^{-1}$ expected for conical jets \citep{blandfordkonigl79}. 

Therefore, if the interday variables from this study and the 5\,GHz scintillators from the MASIV survey are drawn from the same underlying source populations, their source sizes must exhibit a frequency dependence of $\theta \propto \nu^{-k}$, where $k$ is between 1.5 to 2. These values of $k$ are in fact consistent with angular broadening due to scattering in the ISM, where the size of the scattering disk or scattered source image, $\theta_{\rm scatt}$, is expected to scale with frequency as $\theta_{\rm scatt} \propto \nu^{-2}$ \citep[e.g.,][]{rickett90}. For example, scatter broadening at a second, more distant screen in the ISM can cause the apparent source size to increase more rapidly with decreasing frequency, compared to the frequency scaling of the intrinsic source size. This leads to the suppression of the scintillation amplitudes at 5\,GHz relative to that at 15\,GHz, as observed through the more nearby scattering screen primarily responsible for the ISS. This 2-screen scattering example is analogous to the suppression of solar-wind induced interplanetary scintillation of compact radio sources, when observed through sight-lines with strong interstellar scattering \citep[e.g.,][]{duffett-smithreadhead76}. Indeed, a previous study has shown that the frequency dependence of ISS amplitudes measured at 5 and 8\,GHz simultaneously is consistent with a $\theta \propto \nu^{-2}$ relation for sources observed through strongly scattered sight-lines \citep{koayetal12}. This explanation is supported by the fact that the fraction of sources exhibiting significant 15\,GHz ISS increases dramatically for sources observed through $I_{\alpha} \geq 10\,$R compared to that with $I_{\alpha} < 10\,$R.

Another simpler explanation for the relatively high amplitude ISS at 15\,GHz is that these interday variables are observed through sight-lines where the transition frequency between weak and strong ISS is about $\nu_{0} \sim 15 \,$GHz or higher. This implies that at 15\,GHz, the sources are scintillating in the strong scattering regime (or close to the boundary between weak and strong scattering), as opposed to weak ISS as assumed above. Assuming $\nu_{0} \sim 15 \,$GHz, we estimate from the fitting function that compact components of $\theta \sim 20$ to $100\,\mu$as will exhibit ISS of 3 to 13\% modulation indices at an observing frequency of 15\,GHz, comparable to our observations. Assuming similar underlying source populations between the OVRO and MASIV Survey samples, the source size-frequency scaling would be more consistent with that expected for conical jets. 

\subsection{Interday variability of Galactic-plane blazars}
\label{galacticplaneblazars}

Sources observed at low Galactic latitudes have been found to exhibit rapid variations at cm wavelengths \citep{taylorgregory83}, attributed to refractive ISS \citep{rickett86}. These include the extragalactic source CL4 \citep{keenetal73,margonetal81}, which has been observed to exhibit variability on a timescale of weeks at 5\,GHz and 15\,GHz \citep{websterryle76}. In fact, \citet{seaquistgilmore82} report 15\,GHz interday flux variations in CL4, likely ISS caused by enhanced scattering at the Cygnus Loop.

While our CGRaBS-selected sample does not include sources at Galactic latitudes $\lvert b \rvert < 10^{\circ}$, the radio counterparts of two Fermi-detected Galactic plane blazars, 3EG\,J2016$+$3657 and 3EG\,J2027$+$3429, show clear visual evidence of interday variability in their 15\,GHz lightcurves \citep{karaetal12}, and have also been monitored by OVRO since 2008. 3EG\,J2027$+$3429 (J2025$+$3343) even appears to exhibit an ESE \citep{karaetal12,pushkarevetal13}. Although these two sources are not a part of our sample, we examine their variability amplitudes here and compare them with that of our sample.

From the OVRO lightcurves, we derived $D(\rm 4d)$ for these two sources and obtain $m_{D(\rm 4d)}$ of 3\% and 5\% respectively. Their $m_{D(\rm 4d)}/m_{\sigma}$ values are 1.72 for 3EG\,J2016$+$3657 and 2.76 for 3EG\,J2027$+$3429, so the latter would have been selected as an interday variable based on our selection criteria. From Figure~\ref{mvsS15}, we can clearly see that these two sources, shown as black squares, are among the most variable of the strong $> 1$\,Jy level sources. 

At Galactic latitudes of $\lvert b \rvert < 3^{\circ}$, these two sources are observed through a highly turbulent ISM and heavily scattered sight-lines in the direction of the Cygnus OB1 association \citep{spanglercordes98}, with $I_{\alpha}$ of 88.4\,R and 29.5\,R for 3EG\,J2016$+$3657 and 3EG\,J2027$+$3429 respectively. Their large 4-day variability amplitudes at 15\,GHz further strenghten our argument that ISS is responsible for the interday varability of these two sources and the most variable blazars in our CGRaBS sample.

\subsection{Intermittent scintillators}\label{intermittent}

As mentioned in Section~\ref{variables}, of the 53 sources that are observed through sight-lines of $I_{\alpha} \geq 10 \rm \, R$, only 11 of them were selected as interday variables. One possible explanation for this is that the WHAM $I_{\alpha}$ measurements were obtained at an angular resolution of 1$^{\circ}$, much larger than the typical tens to hundreds of micro-arcsecond source sizes of scintillating components in blazars. The high values of $I_{\alpha}$ may not be representative of the actual sight-line towards the source.

The strength of ISS is not only dependent on the line-of-sight scattering strength, but also on the compactness of the source \citep{narayan92, rickett90, koayetal18}. The non-scintillating sources seen through strongly scattered sight-lines may simply not be sufficiently compact to exhibit significant ISS. 

Additionally, the most compact components in these weakly variable sources may also be transient and not persistent over the entire 10-year observing span of the OVRO monitoring program. Our selection criteria for the interday variables is biased towards sources that exhibit persistent variability on these short timescales over a significant portion of the full 10 year observing span. For sources with intermittent ISS, the mean variability amplitudes that we measure over the full 10 years will be suppressed by the low variability amplitudes during epochs when the source is not scintillating. \citet{lovelletal08} found that only 25\% of flat-spectrum extragalactic sources scintillate in either 3 or all 4 epochs of the 5\,GHz MASIV Survey, i.e. are persistent scintillators. Interestingly, this fraction is consistent with the fraction (21\%) of interday variables in the sample of sources with high H$\alpha$ intensities ($I_{\alpha} \geq 10 \rm \, R$).  

The lightcurve of J0502$+$1338 (Figure~\ref{J0502}) illustrates the potential effect of ISS intermittency on the selection of variables \citep[][]{jaunceyetal19}. When the source is in a low state (with low mean flux densities), the amplitude of interday variability is relatively low. When the flux density increases, possibly due to a flare, the interday variability increases in amplitude. This can be physically explained by the ejection of a compact, scintillating component during the flare. The ISS may persist until the compact component expands and dissipates. The line-of-sight H$\alpha$ intensity of J0502$+$1338 is 12.7\,R, and it is one of the sources observed through the Orion-Eridanus superbubble (see Section~\ref{orionob1} below). But its $m_{D(\rm 4d)}/m_{\sigma}$ of 1.4 is below our selection threshold of 2; it was thus not selected as a significant interday variable source due to the fact that it scintillates strongly during only half of the observing period. 

Besides changes in intrinsic source compactness, inhomogeneities in the structure of the intervening scattering screen can also cause intermittent ISS \citep[][]{kedziora-chudczer06, koayetal11b, debruynmacquart15, liuetal15}, again resulting in lower mean variability amplitudes measured over the entire OVRO observing span. 

The fraction of sources that exhibit significant ISS at one time or another is thus likely to be larger than the fraction of the most significant interday variables we identified, when these intermittent scintillators are included. For example, there are 77 sources in our sample with $m_{D(\rm 4d)} \geq 1.5m_{\sigma}$. Of these, 21 of them are observed through sight-lines with $I_{\alpha} \geq 10 \rm \, R$, constituting 40\% of the high $I_{\alpha}$ sample. On the other hand, only 5\% of the $I_{\alpha} < 10 \rm \, R$ sample exhibit these $> 1.5m_{\sigma}$ variations. This example not only demonstrates the robustness of our result regardless of the selection threshold for the most significant variables, but also that ISS is still present in sources whose variability amplitudes are less significant. 

Finding and confirming more of these sources will enable us to examine if this intermittency is mainly due to changes in source structure or the intervening ISM. The former may cause an increase in IDV during flaring states in blazars, as seen in J0502$+$1338 (Figure~\ref{J0502}). Based on a visual inspection of all sources with $I_{\alpha} \geq 10$\,R, other intermittent scintillator candidates include J0559$-$1817, J0619$-$1140, J0630$-$1323, J1617$-$1122, and J1619$-$1817. More sophisticated methods are required to systematically search for such intermittent scintillation in the OVRO data. One possibility is to separate each lightcurve into multiple epochs, and to calculate $D(\rm 4d)$ in each epoch separately or only during the high flux density states of the sources. This is beyond the scope of the present paper, and will be explored in follow-up studies.

\subsection{Intrinsic variability vs ISS of individual sources}\label{intrinsicvsISS}

While the Galactic dependence of the variability amplitudes confirms at a statistical level the contribution of ISS to the 15\,GHz interday variability of a significant fraction of the interday variables and our full sample of sources, we cannot ascertain the origin of the interday variability of an individual source based solely on observations at a single frequency. 

For example, since a flare can also lead to fast-timescale intrinsic variability due to the compactness of the new source component, the large flux density variations observed in J0502$+$1338 (Figure~\ref{J0502}) during its high state may also have an intrinsic origin. Of the nine interday variables that have line-of-sight $I_{\alpha} < 10$\,R, six of them, notably J0721$+$7120 and J1135$-$0428, exhibit large flares on timescales less than a year (Figure~\ref{variableslcsf}), which may have skewed their 4-day variability amplitudes towards larger values. However, we argue that a component that is compact enough to exhibit intrinsic variability on interday to monthly timescales must also be sufficiently compact enough to scintillate, if the line-of-sight is highly scattered \citep{koayetal18}, which is clearly the case for J0502$+$1338. 

J0721$+$7120 (S5 0716+714) is in fact well-known as a highly varable source at radio, optical, X-ray and gamma-ray wavelengths \citep[e.g.,][]{fuhrmannetal08, guptaetal12}. Intra and interday variability has been detected for this source at cm and mm wavelengths \citep{agudoetal06,guptaetal12,leeetal16}. Although the 5\,GHz intraday variations appear to exhibit annual cycles \citep{liuetal12,liuetal13}, characteristic of ISS, the origin of the intraday variations observed between 10 to 15\,GHz is still strongly debated \citep{jaunceyetal19}. This highlights the complexity in distinguishing between ISS and intrinsic variability in individual sources.

\subsection{The candidate extreme scintillator J0616$-$1041}\label{extremeISS}

J0616$-$1041 (Figure~\ref{J0616} in Appendix~\ref{appendixvariables}) exhibits large 18\% flux density variations on a timescale of 4 days. Such high 4-day variability amplitudes, the strongest in our entire sample of interday variables after excluding the problematic source J0259$-$0018 (Appendix~\ref{j0259}), is almost comparable to that exhibited by the so-called `extreme scintillators', of which only a handful are known, including PKS0405$-$385 \citep{kedziora-chudczeretal97}, J1819$+$3845 \citep{dennett-thorpedebruyn00}, and PKS1257$-$326 \citep{bignalletal03}. 

The large amplitude variations observed in J0616$-$1041 cannot be attributed to errors in flux density measurements alone, even though flux-independent errors, such as thermal noise, are more significant for a weaker source. For the flux measurement errors of $m_{\sigma} \sim 5.5\%$ for J0616$-$1041, equivalent to $D_{\sigma} = 2m_{\sigma}^2 = 0.006$, assuming that the noise is white and independent of time lag, $D_{\sigma}$ contributes additively to the measured $D(\tau)$ across all values of $\tau$. This will increase the measured amplitude of flux density variations. However, even if we subtract $D_{\sigma} = 0.006$ from $D(\rm 4d) = 6.32 \times 10^{-2}$ to account for the noise contribution, the resultant $m_{D(\rm 4d)}$ is still $\sim 17\%$. Furthermore, the $m_\sigma$ of J0616$-$1041 is comparable to that of the other weak $\sim$0.2\,Jy sources in our sample (Figure~\ref{mvsS15}). It is therefore unlikely that the $m_\sigma$ of J0616$-$1041 is underestimated.

If the large amplitude interday variability of J0616$-$1041 is indeed due to ISS, the detection of 1 extreme scintillator in our sample of 1158 sources is statistically consistent with the non-detection of any new extreme scintillators in the MASIV Survey sample of $\sim$500 sources \citep{lovelletal08}.

Such extreme scintillation can occur if the source is ultra-compact, or if there is an intervening, highly turbulent scattering screen located relatively close to the Earth. With a line-of-sight $I_{\alpha} = 18.94$\,R, J0616$-$1041 appears to be observed through a heavily scattered sight-line. Assuming that the transition frequency between weak and strong scintillation is $\nu_{0} = 15$\,GHz at the line-of-sight towards this source, as given by \citet{walker01}, we estimate that to exhibit such high amplitude scintillation, the source must be about 16 $\mu$as in angular size for a scattering screen located at the typical distance of 500\,pc from the Earth. At a mean flux density of $\sim$0.2\,Jy, such a compact source would have an apparent brightness temperature of $\sim 10^{12}$\,K. Assuming an equipartition brightness temperature of $\sim 10^{11}$\,K \citep{readhead94, lahteenmakietal99}, a Doppler boosting factor of $\sim 10$ is required, well within the measured range of values for blazars \citep{hovattaetal09}. For a less stringent source compactness of 100$\mu$as, the scattering screen has to be very close, of order $\sim 10\,$pc away from the Earth. 

For the well-known extreme scintillators such as PKS1257$-$326 and J1819$+$3845, the high amplitude variations are attributed to the presence of nearby ($< 10$\,pc), highly turbulent scattering screens \citep{bignalletal06,macquartdebruyn07,debruynmacquart15,vedanthametal17c}, rather than to the compactness of the sources themselves. Due to the very nearby scattering screens, an important feature of these well-known extreme scintillators is their rapid intraday (and even intrahour) variability timescales. For example, PKS1257$-$326 is known to vary in flux density by up to 40\% on a timescale of 45 minutes \citep{bignalletal03}. Follow-up observations of J0616$-$1041 at a higher intraday cadence are required to confirm its status as a rapid scintillator in the mould of the well-known extreme, intra-hour scintillators like PKS1257$-$326, as well as providing better constraints on the properties of the scattering screen. If the rapid variations are confirmed, metre-wave polarimetry can be used to `image' the scattering cloud towards the source, as was done for J1819$+$3845 \citep{vedanthametal17c}, thus providing strong contraints on its distance and structure. There are, however, hints that the scattering screen of J0616$-$1041 may be a few hundred pc away from the Earth, which we discuss later in Section~\ref{orionob1extreme}.

\subsection{Clustering of interday variables through the Orion-Eridanus superbubble}\label{orionob1}

\begin{figure*}
	\begin{center}
		\includegraphics[width=\textwidth]{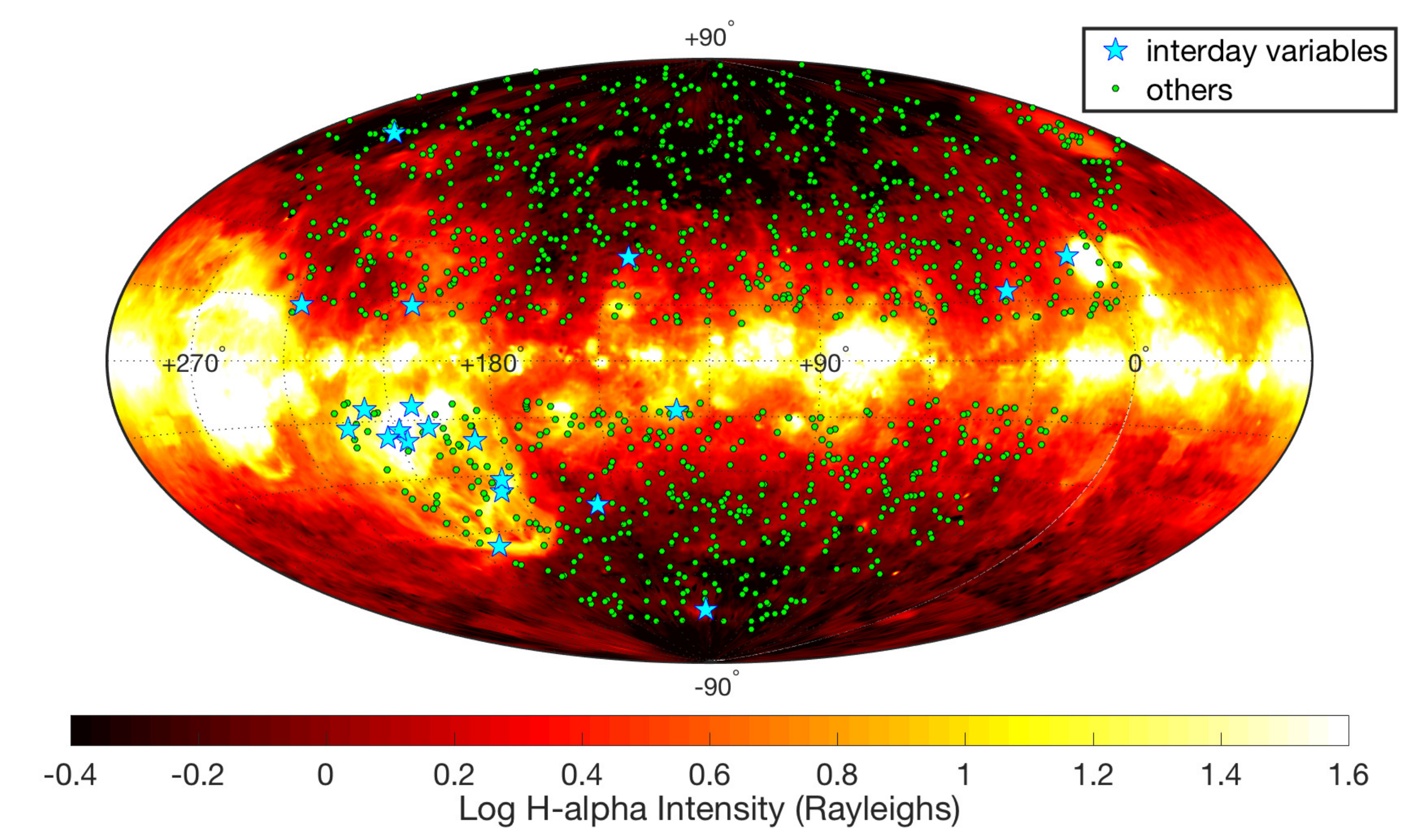}
	\end{center}
	\caption{{All-sky H$\alpha$ intensity map from the WHAM Survey \citep{haffneretal03, haffneretal10} shown in Galactic coordinates. The coordinates of the 1158 sources in the OVRO sample are shown as green circles or blue stars, where the latter represents the 20 sources that exhibit the most significant 4d variability amplitudes ($m_{D({\rm 4d})} \geq 2 m_{\sigma}$, excluding J0259$-$0018). \label{halphamap}}}
\end{figure*}

Figure~\ref{halphamap} shows the distribution of the OVRO blazars on the sky in Galactic coordinates. The interday variables are shown as blue stars. The colour map shows the H$\alpha$ intensities from the WHAM Survey \citep{haffneretal03}, where we also include the data from the Southern sky survey \citep{haffneretal10}. 

The sky distribution of the interday variables shows a clear dependence on the structures of the ionized gas in our Galaxy, strengthening the argument for their ISS-induced variability. In particular, more than half (11 of 20) of the interday variables are observed through the highly ionized region between longitudes of 175$^{\circ}$ to $240^{\circ}$ and latitudes of $-10^{\circ}$ to $-50^{\circ}$. This region is associated with the Orion-Eridanus superbubble, which contains gas with properties similar to that of the warm ionized medium \citep{odelletal11}, photo-ionized by the hot, giant stars of the Orion OB1 association \citep{brownetal94,brownetal95}. 

\subsubsection{Origin of extreme scintillation?}\label{orionob1extreme}

Recently, the scattering structures responsible for the extreme scintillation of PKS1257$-$326 and J1819$+$3845 have been found to be associated with nearby hot stars Alhakim and Vega respectively \citep{walkeretal17}. They appear to be radially elongated filamentary structures pointing towards the host star, which \citet{walkeretal17} suggest to be cometary `tails' of molecular globules analogous to that observed in the Helix nebula. Interestingly, J0616$-$1041 is observed through the Orion-Eridanus superbubble. Perhaps the scattering screen of J0616$-$1041, as well as that of the other IDV sources observed through this region, may comprise similar anisotropic structures associated with the hot O and B stars in the region. If this is the case, the fact that these scattering structures are located at about 300 to 500\,pc away from the Earth \citep{brownetal95}, suggests that J0616$-$1041 may indeed be ultra-compact.

We note that J0616$-$1041 also happens to be the lowest flux density source among the 20 interday variables. If these compact blazars are brightness temperature-limited, either due to the inverse-Compton catastrophe \citep{kellermannpauliny-toth69} or energy equipartition between the magnetic fields and electrons \citep{readhead94, lahteenmakietal99}, their angular sizes are expected to scale as $\theta \propto \sqrt{S_{15}}$. The weakest sources are also therefore most likely to be the most compact in angular size and thereby scintillate more strongly \citep{lovelletal08,koayetal18}.

\subsubsection{Implications for the high-energy neutrino source TXS\,0506$+$056}
\label{txs0506}

The blazar TXS\,0506$+$056 (J0509$+$0541) was recently identified as a source of high energy neutrinos \citep{icecubeetal18a,icecubeetal18b}. It is observed through the Orion-Eridanus superbubble, with $I_{\alpha} = 23.3\,$R. This source is in our sample, but its $m_{D(\rm 4d)}/m_{\sigma}$ of 1.7 means it narrowly also missed being classified as on of the interday variables. It is a well known scintillator at lower frequencies \citep[][Edwards et al. in prep]{lovelletal08}. The fact that this source is observed through this special region in our Galaxy means that any attempt to interpret its radio intra/interday variability \citep{tetarenkoetal17} in connection to that at other wavelengths, or to its intrinsic jet properties, will need to be carried out with caution.

\section{Summary and Conclusions}\label{summary}

In this study, we have characterized the 15\,GHz variability amplitudes of 1158 radio-selected blazars monitored over $\sim$10 years by the OVRO 40-m telescope, at the shortest observed timescale of 4\,days, to determine the origin of the interday flux density variations. Our main findings can be summarized as follows:

\begin{enumerate}
\item The 4 to 20-day structure function amplitudes show a significant dependence on line-of-sight Galactic H$\alpha$ intensities, demonstrating the presence of interstellar scintillation in the OVRO blazar lightcurves on timescales of days and weeks.

\item Of the 1158 sources, we identified 20 that exhibit significant interday variability on 4-day timescales. Based on the higher fraction (21\%) of these interday variables detected through sight-lines with $I_{\alpha} \geq 10$\,R compared to only 0.8\% detected through weakly scattered sight-lines of $I_{\alpha} < 10$\,R, we argue that the 3 to 13\% flux density variations observed in these sources are mainly driven by ISS. 

\item ISS is likely also present in the interday variations of a larger fraction of sources that exhibit less significant variability, i.e., $1.5 \lesssim m_{D(\rm 4d)}/m_{\sigma} < 2$. Our selection of significant variables missed out on intermittent scintillators such as J0502$+$1338 that are observed through heavily scattered sight-lines, but exhibit significant ISS only during high flux density states; we interpret this as due to the ejection of new compact components that scintillate during a flare. Improved methods need to be developed to search for and identify such intermittent scintillators in these long term data, to better understand the full population of scintillating sources at 15\,GHz.

\item We have identified J0616$-$1041, displaying 18\% flux density variations on 4-day timescales, as a candidate extreme scintillator. This source either contains an ultra-compact core of order $\sim 10\mu$as, or is observed through a highly-turbulent scattering screen located no more than 10 parsecs away from the Earth. Follow-up observations will enable us to confirm if this candidate is indeed scintillating rapidly on intraday timescales.

\item Of the 20 sources we classified as interday variables, more than half of them (11 sources, including J0616$-$1041), are observed through the Orion-Eridanus superbubble. This highly turbulent and ionized region appears to be an important region of interstellar scattering at 15\,GHz. The high-energy neutrino source TXS\,0506$+$056 is observed through this region, so its intra/interday radio variability will need to be interpreted with this in mind. 

\end{enumerate}

ISS is already known to dominate the intra and interday variability of compact, flat-spectrum radio sources up to 8\,GHz. While ISS is typically ignored or assumed to be unimportant at 15\,GHz, we have demonstrated through this work that ISS is still a significant contributor to intra and interday variability of compact sources at this frequency, especially through heavily scattered sightlines with high electron column densities, i.e. $I_{\alpha} \geq 10\,R$. These short term ISS-induced flux density variations are often superposed on larger amplitude intrinsic variations occurring on longer timescales of $\gtrsim 100\,$days.

In order to distinguish between ISS-induced and intrinsic inter/intraday variations of AGNs at 15\,GHz and below, coeval monitoring at multiple frequencies, including at above 20 to 40\,GHz, is required. Even then, opacity effects may smear out the rapid intrinsic variations at lower frequencies, making it difficult to search for cross-correlated variability in multi-frequency lightcurves. Future surveys and monitoring of AGN radio variability will therefore need to be conducted at frequencies greater than 15\,GHz if the goal is to study the intrinsic causes of intraday variability in radio AGNs, particularly if the sources are seen through thicker regions of the Galaxy. An alternative is to select sources only at higher Galactic latitudes where ISS at 15\,GHz is less significant.

\section*{Acknowledgements}

We thank the anonymous reviewer for the helpful suggestions to improve the manuscript. J.Y.K. thanks Keiichi Asada, Satoki Matsushita, Wen-Ping Lo, and Geoff Bower for helpful discussions. T.H. was supported by the Academy of Finland projects \#317383 and \#320085. W.M. and R.R. acknowledge support from CONICYT project Basal AFB-170002. This research has made use of data from the OVRO 40-m monitoring program (Richards, J. L. et al. 2011, ApJS, 194, 29) which is supported in part by NASA grants NNX08AW31G, NNX11A043G, and NNX14AQ89G and NSF grants AST-0808050 and AST-1109911. The Wisconsin H$\alpha$ Mapper and its H$\alpha$ Sky Survey have been funded primarily by the National Science Foundation. The facility was designed and built with the help of the University of Wisconsin Graduate School, Physical Sciences Lab, and Space Astronomy Lab. NOAO staff at Kitt Peak and Cerro Tololo provided on-site support for its remote operation.








\appendix

\section{Estimation of flux calibration uncertainties}\label{fluxcalibrationerrors}

\begin{figure}
	\begin{center}
		\includegraphics[width=\columnwidth]{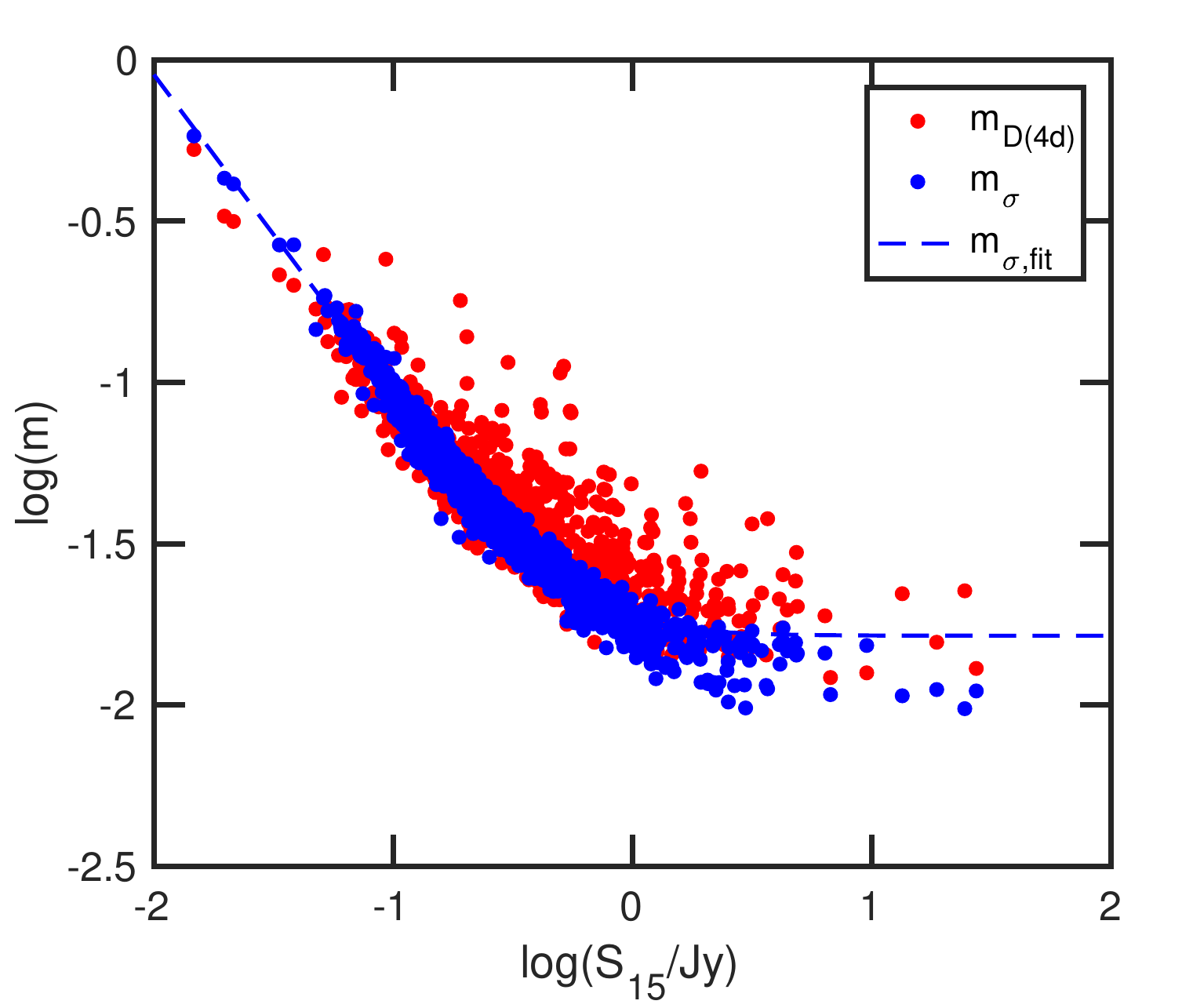}
	\end{center}
	\caption{{Modulation indices derived from the 4-day structure function amplitude, $m_{D(\rm 4d)}$, and $m_{\sigma} = \rm median(\sigma_{\rm err})/S_{\rm 15}$, vs. the 15\,GHz flux density. Flux calibration errors are not included in the estimation of $m_{\sigma}$. \label{mvsS15nofc}}}
\end{figure}

\begin{figure}
	\begin{center}
		\includegraphics[width=\columnwidth]{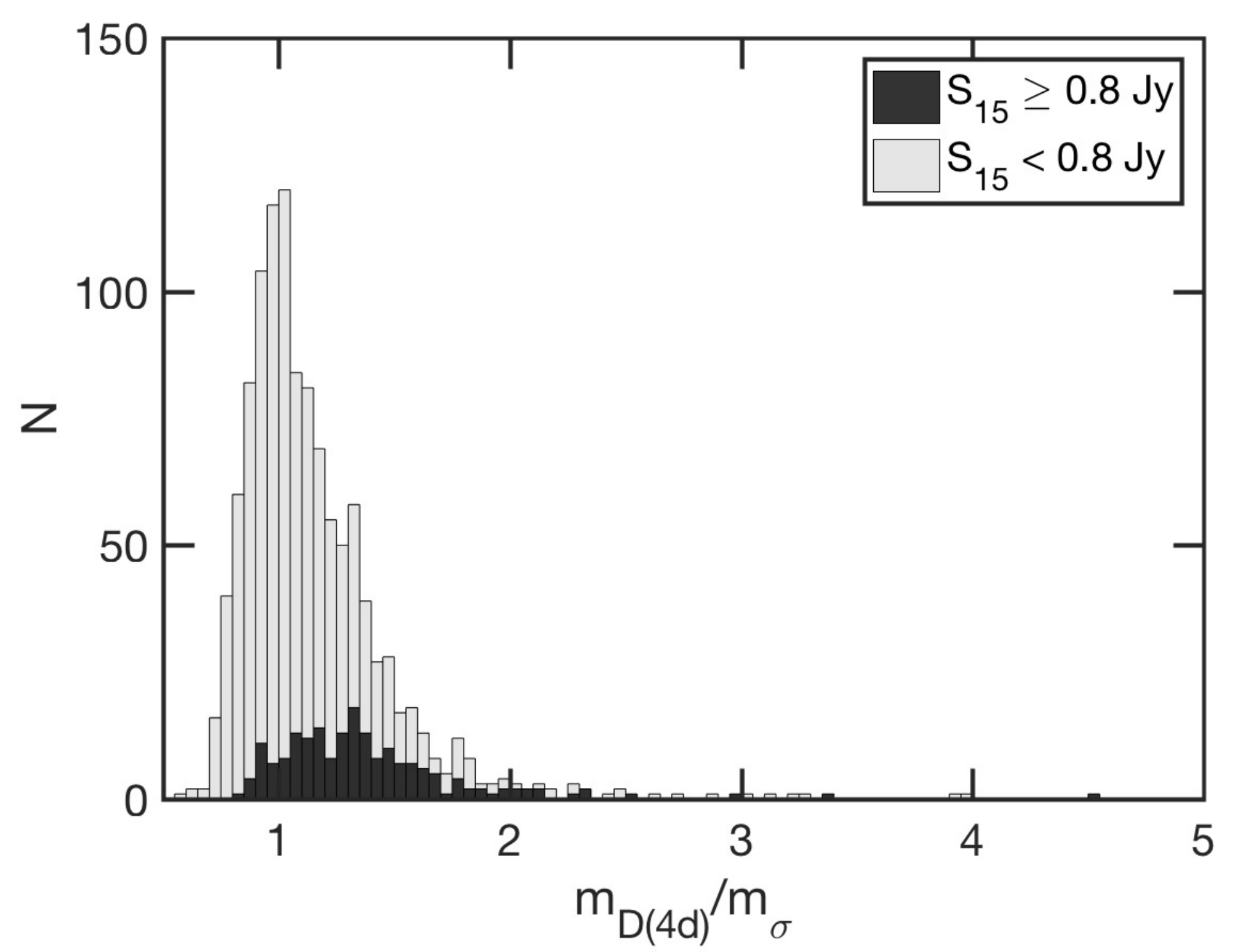}
	\end{center}
	\caption{{Histogram showing the distribution of the ratio of the 4-day variability amplitudes to the flux normalized measurement uncertainties of each source, $m_{D({\rm 4d})}/m_{\sigma}$, where $m_{\sigma} = \rm median(\sigma_{\rm err})/S_{\rm 15}$. Since the flux calibration errors are not included in the estimation of $m_{\sigma}$, it can be seen that the uncertainties are underestimated for the strong sources with $S_{\rm 15} \geq 0.8$\,Jy, for which the $m_{D({\rm 4d})}/m_{\sigma}$ distribution peaks at a value of about 1.3. \label{diffmnoisenofc}}}
\end{figure}

If we consider only $\sigma_{\rm err}$ (Equation~\ref{erroreqrichards}) in estimating the contribution of noise and instrumental effects ($m_{\sigma}$) to the interday variability amplitudes, such that $m_{\sigma} = \rm median(\sigma_{\rm err})/S_{\rm 15}$, we find that this underestimates the $m_{\sigma}$ of the strong sources. This can be seen in Figures~\ref{mvsS15nofc} and \ref{diffmnoisenofc}, where the $m_{D({\rm 4d})}/m_{\sigma}$ peaks at a value higher than 1 for sources above 0.8\,Jy. This suggests that the flux-dependent errors are underestimated. This can be attributed to errors in flux calibration that were not included during the estimation of $\sigma_{\rm err}$ in the OVRO data.

While a flux calibration error of 5\% is typically assumed based on the long term variations observed in the flux calibrators, this is expected to be lower at interday timescales. To constrain the 4-day flux calibration errors, we first obtained the 4-day structure function amplitudes of the flux calibrators 3C286, and DR21. From Equation~\ref{m2D}, we derive their 4-day modulation indices, $m_{D({\rm 4d})}$, to be 1.5\% and 1.3\% respectively. Since these are bright Jy level sources, the 4-day variability amplitudes will be dominated by flux-dependent errors, in addition to any intrinsic source variability that may introduce flux calibration errors to the science targets. These values of $m_{D({\rm 4d})}$ thus provide an upper limit on the the flux calibration errors.

To estimate the flux calibration uncertainties, we let $m_{\sigma}$ of each source be the flux normalized quadratic sum of $\sigma_{\rm err}$ of the source and the flux calibration error:
\begin{equation}
m_{\sigma} = \dfrac{\sqrt{({\rm median}(\sigma_{\rm err}))^2 + (\sigma_{\rm cal})^2}}{S_{\rm 15}}.  \label{msigmawithfceq}
\end{equation}
where the flux calibration error $\sigma_{\rm cal}$ is some fraction of the source flux density. Based on Equation~\ref{msigmawithfceq}, we determined that letting $\sigma_{\rm cal} \approx 1\%$ of the source mean flux density is adequate to shift the peak of the $m_{D({\rm 4d})}/m_{\sigma}$ distribution of the strong sources to unity, as can be seen in Figures~\ref {mvsS15} and \ref{diffmnoise}. This assumes that $m_{D({\rm 4d})}$ should be dominated by noise, instrumental and systematic uncertainties as described in Equation~\ref{msigmaeq}. This value of $\sigma_{\rm cal}$ is also consistent with the upper limits derived from the $m_{D({\rm 4d})}$ of the flux calibrators. We therefore adopt $\sigma_{\rm cal} = 0.01S_{\rm 15}$ for this work, to correct for the underestimation of $m_{\sigma}$ in the strong sources. While we attribute this mainly to flux calibration errors, this term also folds in any residual flux-dependent errors that may be unaccounted for in the $\sigma_{\rm err}$ of the OVRO data. 

\section{Origin of large interday flux variations in J0259$-$0018}\label{j0259}

\begin{figure}
	\begin{center}
		\includegraphics[width=\columnwidth]{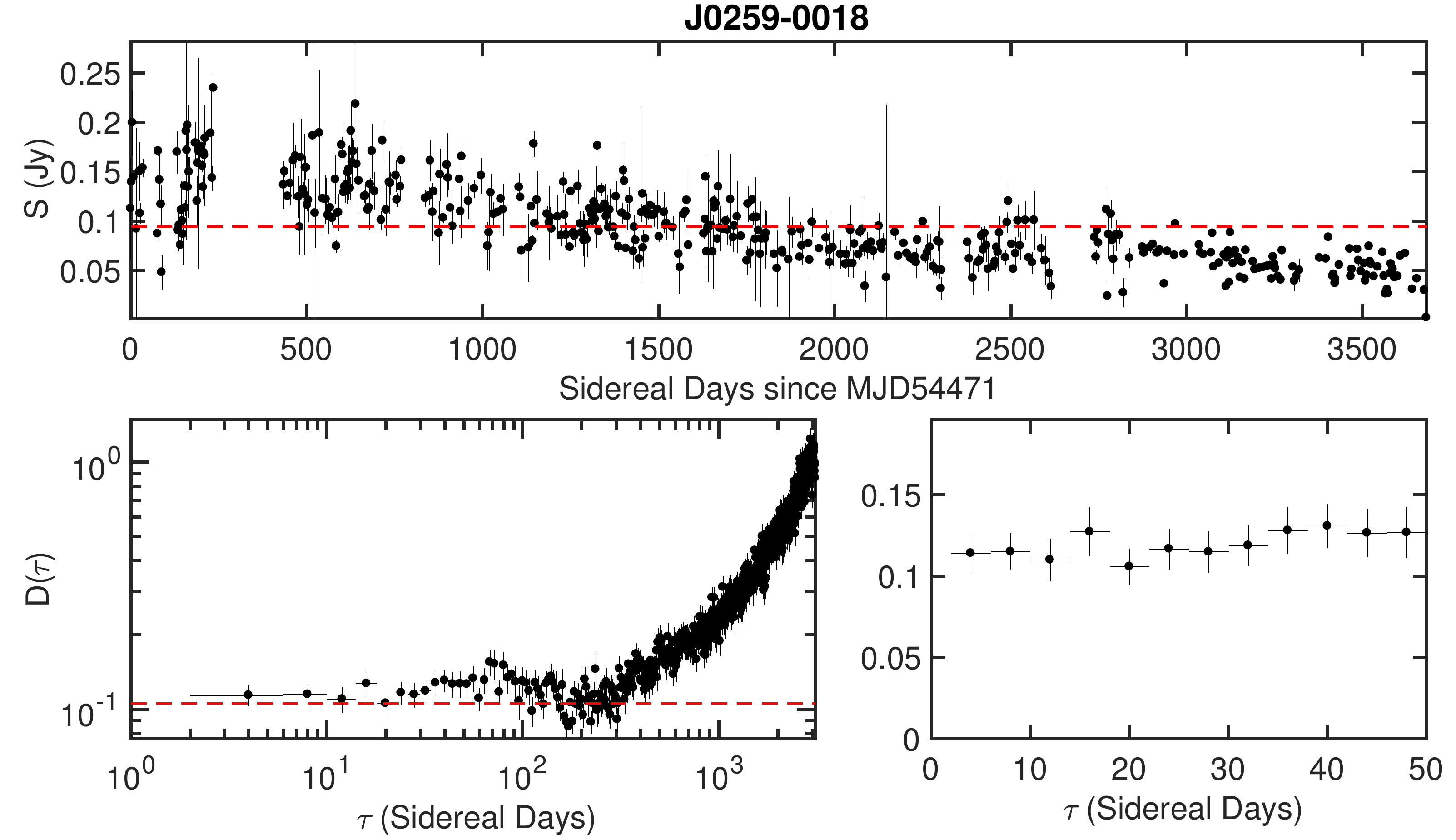}
	\end{center}
	\caption{{Top: Lightcurve for J0259$-$0018, where the horizontal dashed line denotes the mean flux density of the source. The error bars are given by Equation~\ref{erroreqrichards} \citep{richardsetal11}. Bottom: Structure function, $D(\tau)$, shown in its entirety in the left panel, and for $\tau \leq 50 \rm d$ in the right panel. The horizontal dashed line denotes $D_{m15}$ (Equation~\ref{m2D}) derived from the intrinsic modulation indices estimated by \citet{richardsetal14}. \label{J0259}}}
\end{figure}

J0259$-$0018 exhibits 24\% flux density variations on 4-day timescales. Its lightcurve is shown in Figure~\ref{J0259}. It would have been remarkable if these large flux density variations are caused by ISS, as it would be comparable to that observed in a rare class of `extreme scintillators', of which only a handful are known to date \citep{kedziora-chudczeretal97,dennett-thorpedebruyn00,bignalletal03}.

To confirm if the flux density variations observed in J0259$-$0018 have an astrophysical origin, we checked the lightcurves of 4 other sources located within $\pm 5^{\circ}$ of J0259$-$0018 on the sky. These sources are likely to share the same pointing source, and were observed at similar elevations and azimuths at the telescope. Our visual inspections find no correlated interday variability among these sources, so these variations are unlikely to be dominated by pointing or residual gain calibration errors. In any case, such errors are expected to dominate for stronger sources rather than weak ones like J0259$-$0018. Of these 4 nearby sources, 2 of them, J0305$+$0523 and J0318$-$0029, have comparable $\sim$0.1\,Jy mean flux densities to J0259$-$0018. RFI characteristics are expected to be direction dependent, but would equally affect these 2 sources, so cannot explain the excess interday variability observed in J0259$-$001. The $m_{\sigma}$ of J0305$+$0523 and J0318$-$0029 are also comparable to that of J0259$-$0018, but their $m_{D(\rm 4d)}$ are 2 to 4 factors lower than that of J0259$-$0018.

Finally, we checked the VLA Faint Images of the Radio Sky at Twenty-Centimeters (FIRST) Survey \citep{beckeretal95} catalogue to determine if confusion by a nearby bright source could be responsible for the large flux density variations. To our surprise, we found no source detected at the coordinates of RA $=$ 02h59m28.5100s and DEC $= -$00d18\arcmin00.000\arcsec as specified in the CGRaBS and OVRO catalogues for J0259$-$0018. On the other hand, there is a 0.2\,Jy source located exactly $2\arcmin$ South. Checking the VLBA calibrator list, we found this source as J0259$-$0019, with coordinates of RA $=$ 02h59m28.5153s and DEC $= -$00d19\arcmin59.968\arcsec. The flux density variations in the lightcurve of J0259$-$0018 could be caused by the source shifting around within the $2.6\arcmin$ primary beam at different hour angles.

Unless its spectral index is highly inverted such that it is not detectable at 21\,cm at the 0.121\,mJy noise threshold of the FIRST Survey, which is highly unlikely, J0259$-$0018 probably does not exist, and in the original CGRaBS catalogue may in fact be a misidentification of J0259$-$0019. Even if J0259$-$0018 is detectable at 15\,GHz, a source of comparable flux density located $2\arcmin$ away would still lead to confusion and increased flux density variations. We therefore rule out extreme scintillation in J0259$-$0018 and remove it from our list of significant interday variables.

\section{Lightcurves and structure functions of significant interday variables}\label{appendixvariables}

In Figure~\ref{variableslcsf}, we present the 15\,GHz lightcurves measured by the OVRO 40-m telescope for the 20 sources which we detected to be significantly variable (see Section~\ref{variables}), together with their corresponding structure functions which we derived (Section~\ref{sf}). We exclude J0259$-$0018 due to problems described in Appendix~\ref{j0259}.

\begin{figure*}
	\centering
	\begin{subfigure}[b]{.50\linewidth}
		\centering
		\includegraphics[width=.99\textwidth]{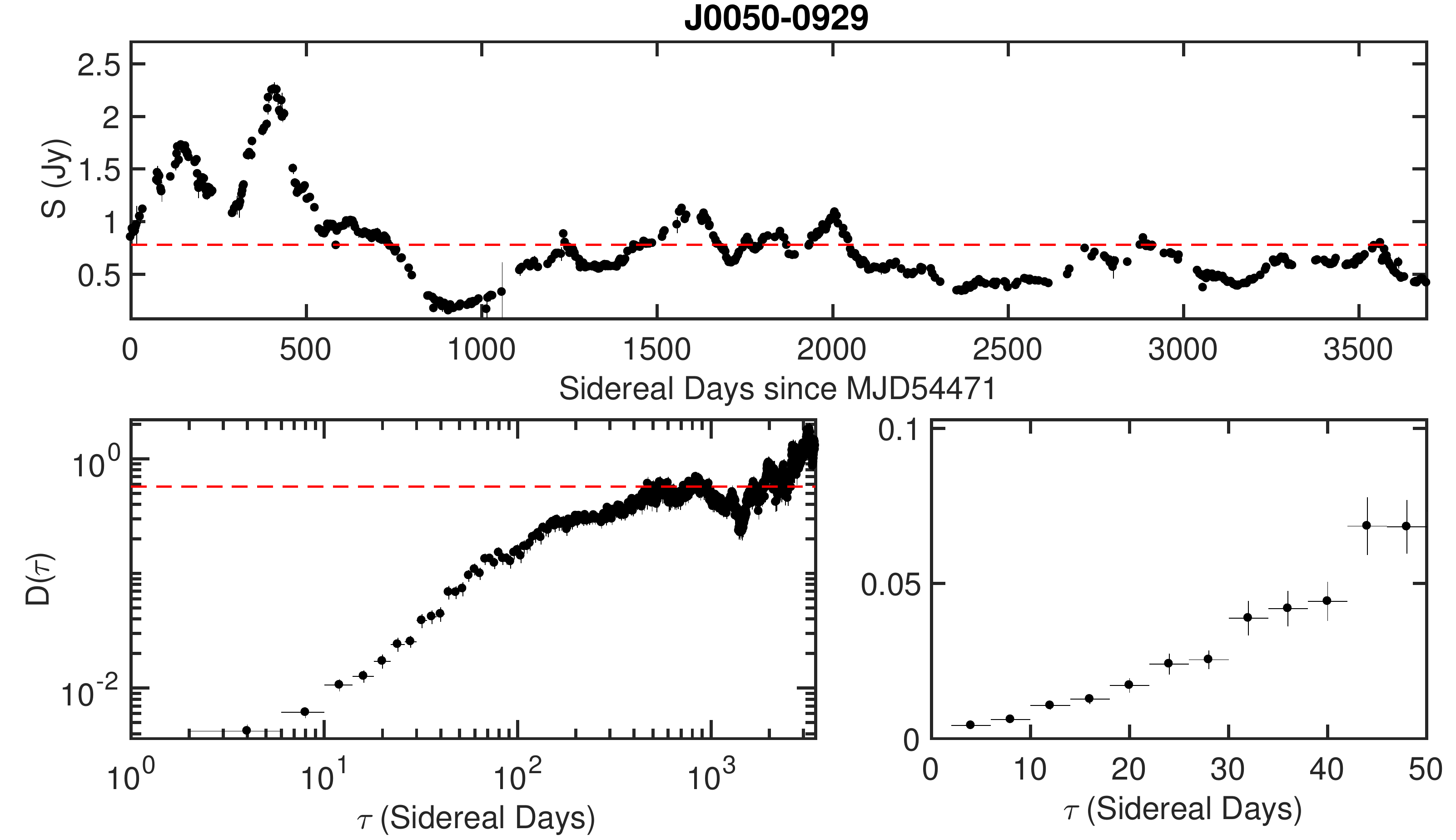}
		\caption{}
	\end{subfigure}%
	\begin{subfigure}[b]{.50\linewidth}
		\centering
		\includegraphics[width=.99\textwidth]{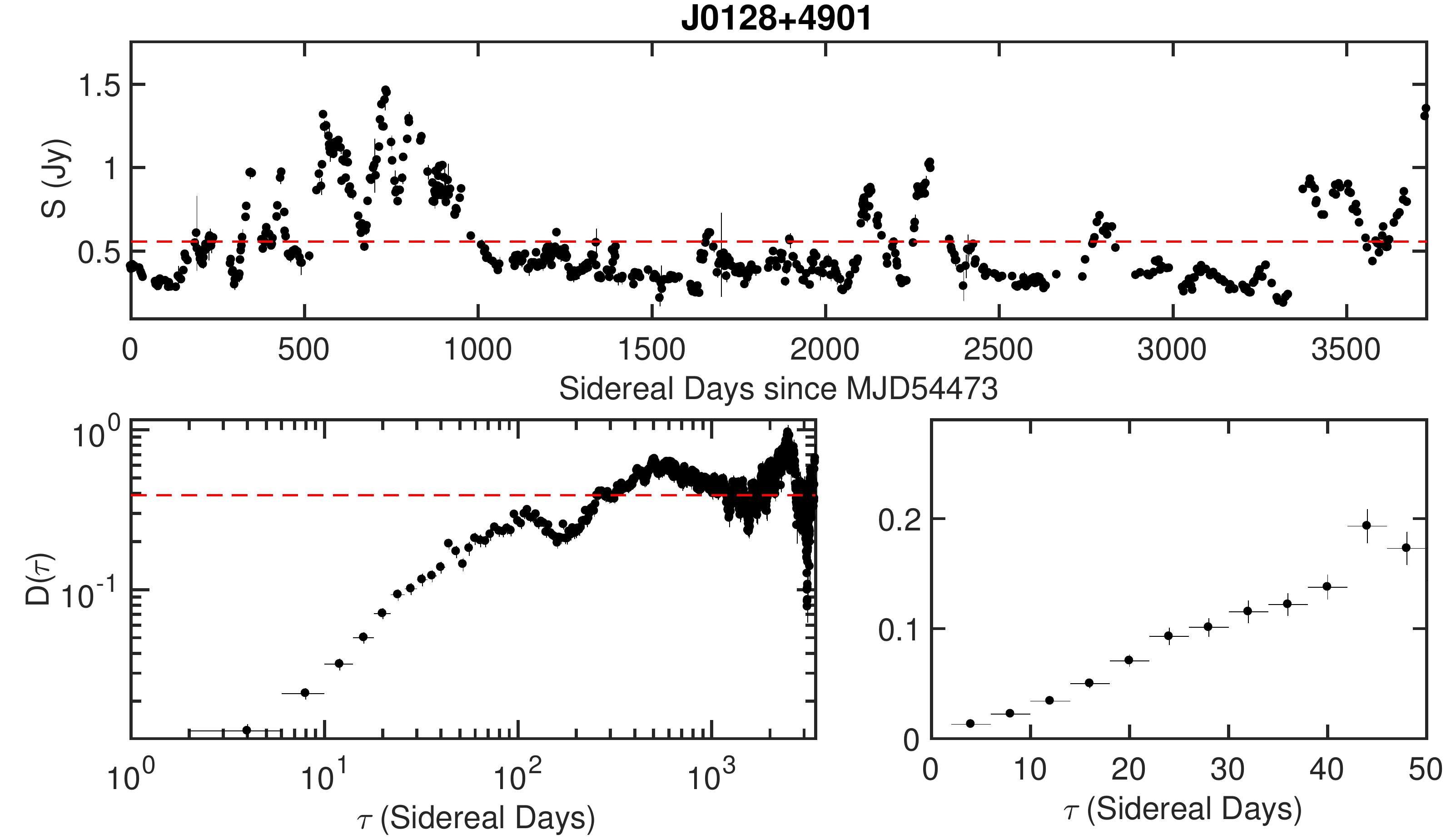}
		\caption{}
	\end{subfigure}\\%
	\begin{subfigure}[b]{.50\linewidth}
		\centering
		\includegraphics[width=.99\textwidth]{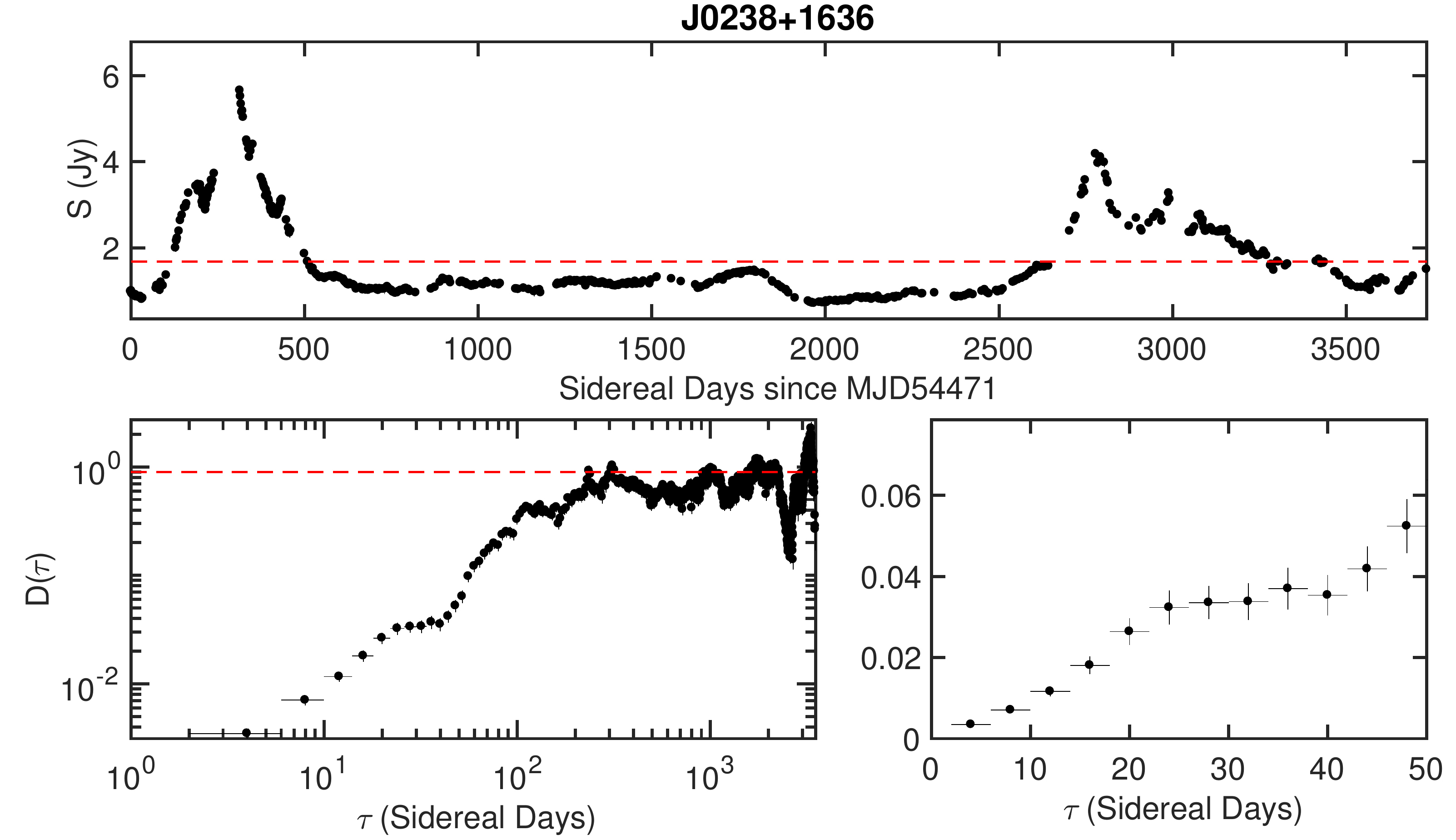}
		\caption{}
	\end{subfigure}%
	\begin{subfigure}[b]{.50\linewidth}
		\centering
		\includegraphics[width=.99\textwidth]{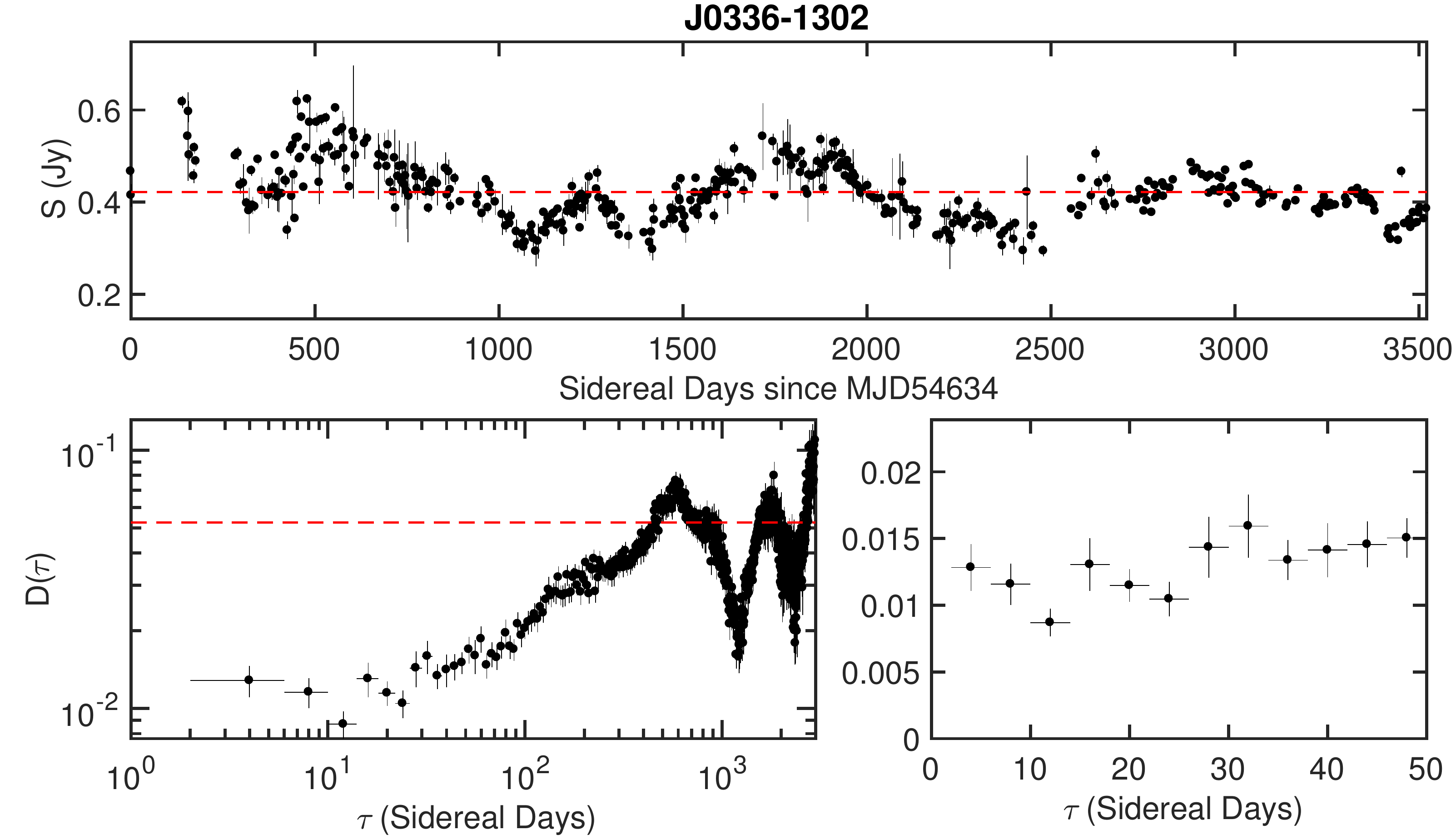}
		\caption{}
	\end{subfigure}\\%

	\begin{subfigure}[b]{.50\linewidth}
		\centering
		\includegraphics[width=.99\textwidth]{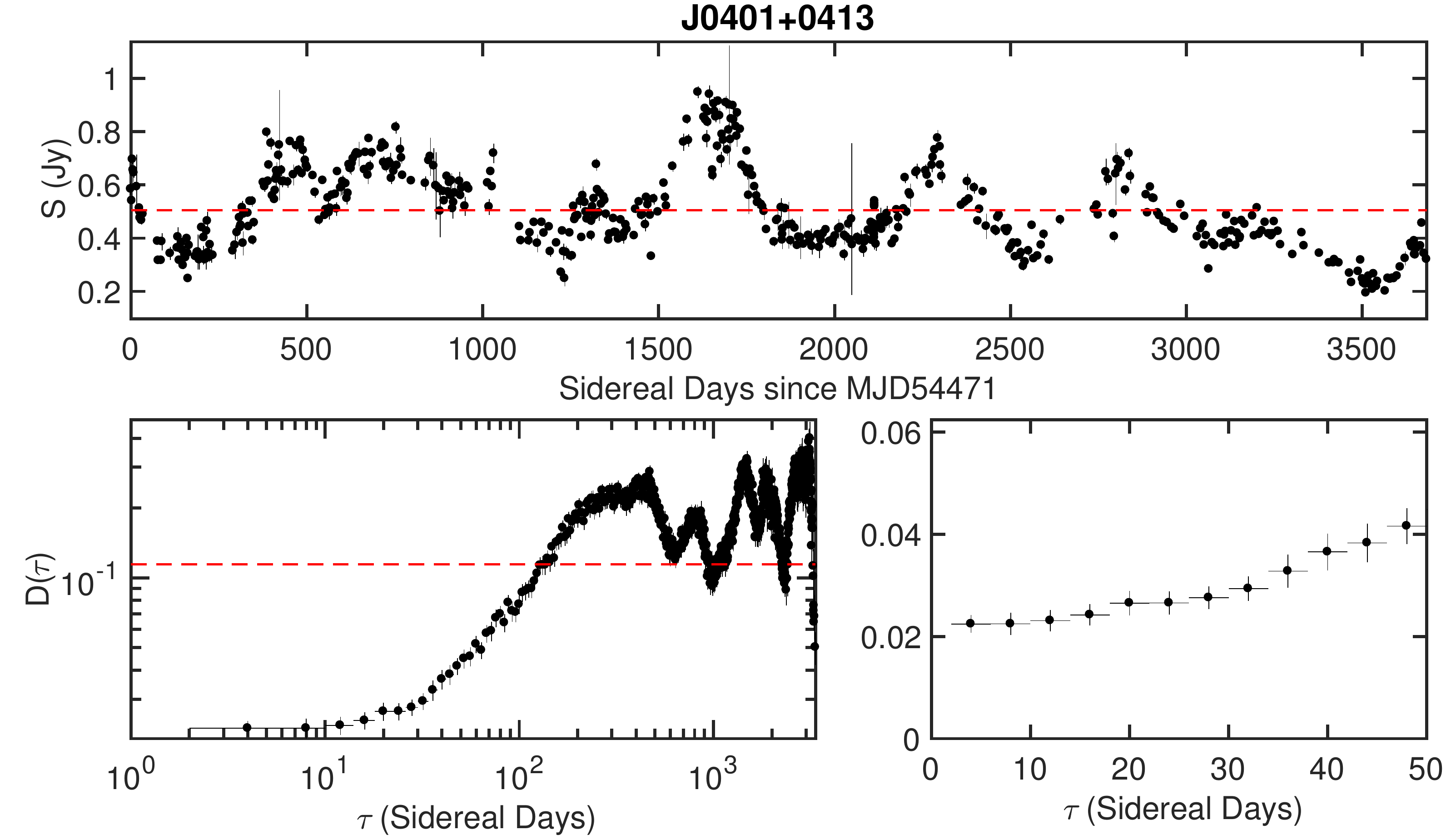}
		\caption{}
	\end{subfigure}%
	\begin{subfigure}[b]{.50\linewidth}
	   \centering
	   \includegraphics[width=.99\textwidth]{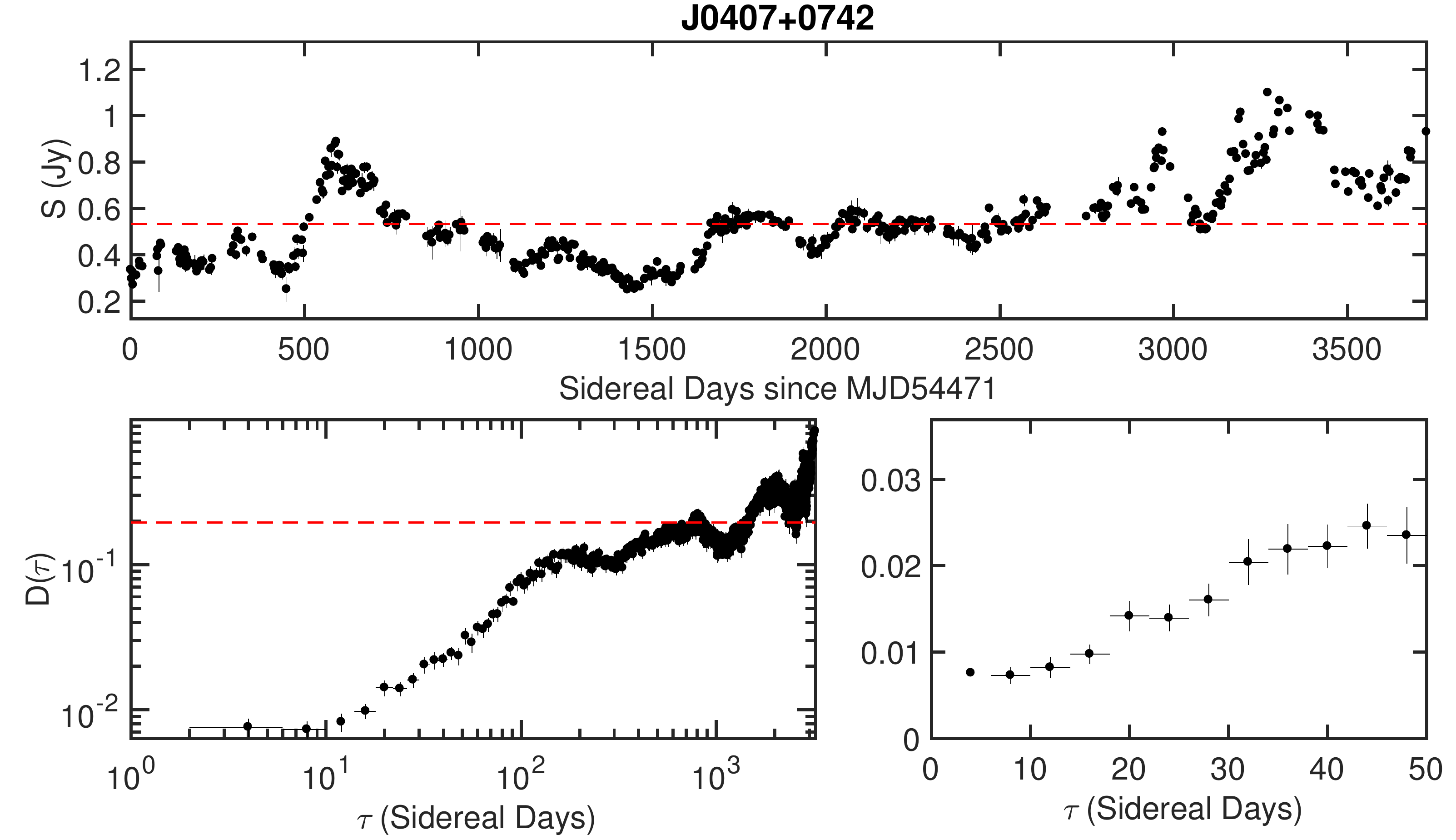}
     	\caption{}
     \end{subfigure}\\%
     \begin{subfigure}[b]{.50\linewidth}
	   \centering
	   \includegraphics[width=.99\textwidth]{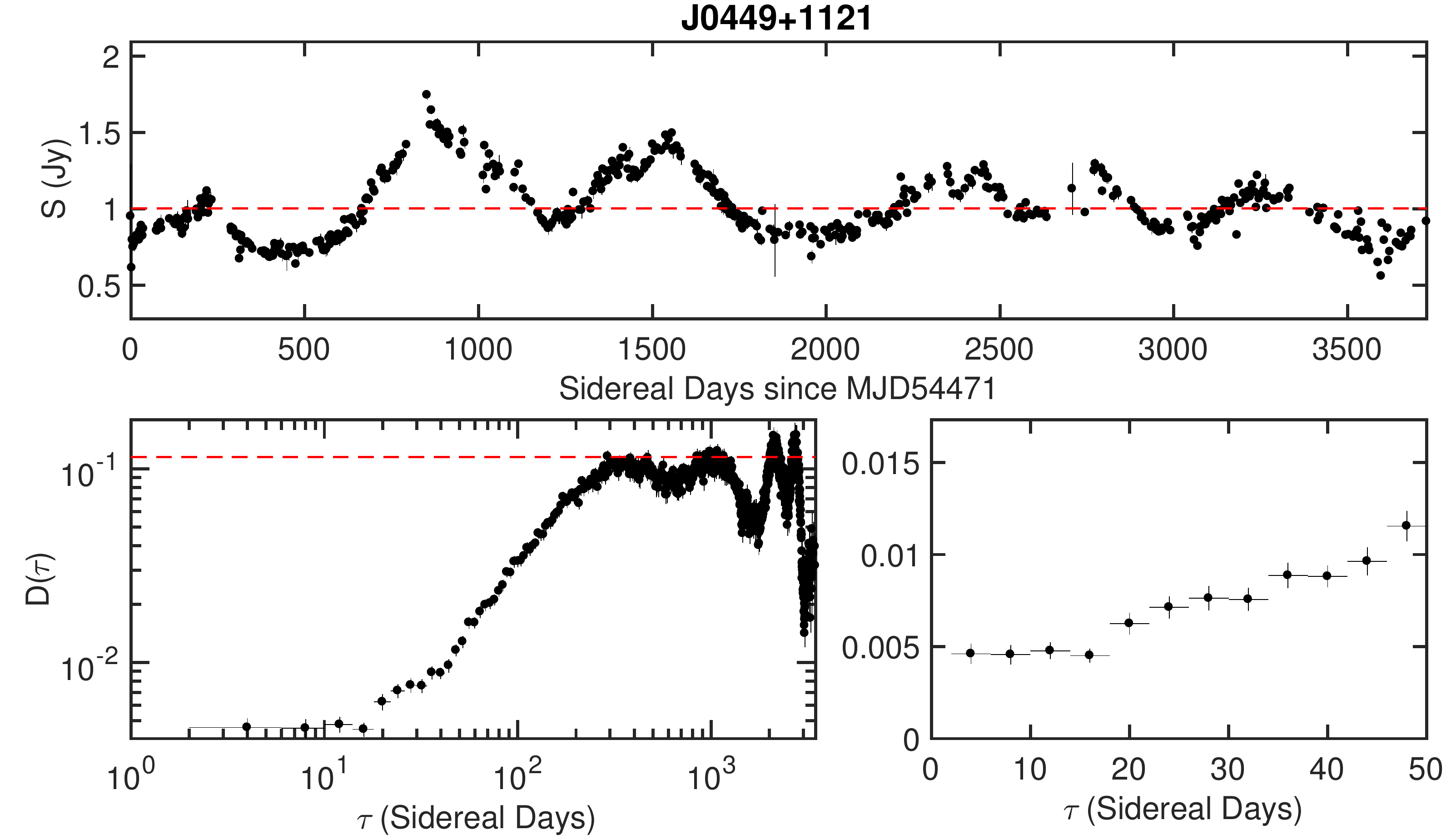}
	   \caption{}
    \end{subfigure}%
    \begin{subfigure}[b]{.50\linewidth}
		\centering
		\includegraphics[width=.99\textwidth]{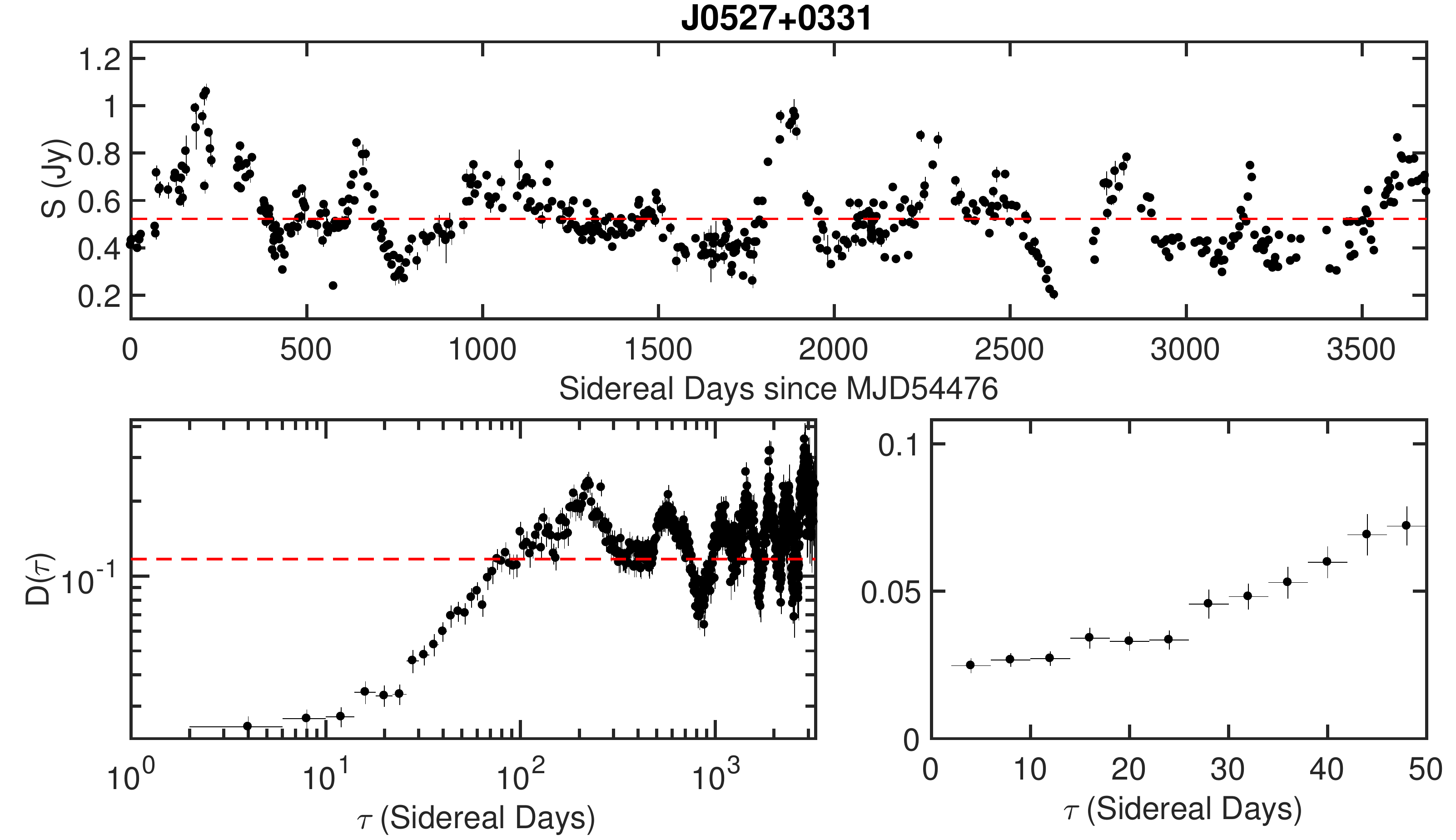}
		\caption{}
	\end{subfigure}\\%
	\caption{Lightcurves and structure functions of the 20 sources exhibiting significant variability amplitudes on 4-day timescales, excluding J0259$-$0018. The lightcurves are shown in the top panel of each subfigure, where the horizontal dashed line denotes the mean flux density of the source. The error bars are given by Equation~\ref{erroreqrichards} \citep{richardsetal11}. The bottom panels of each subfigure show the structure function, $D(\tau)$, in its entirety in the left panel, and for $\tau \leq 50 \rm d$ in the right panel. The horizontal dashed line denotes $D_{m15}$ (Equation~\ref{m2D}) derived from the intrinsic modulation indices estimated by \citet{richardsetal14}.}\label{variableslcsf}
\end{figure*}

\begin{figure*}\ContinuedFloat
	\centering 
	  
	\begin{subfigure}[b]{.50\linewidth}
		\centering
		\includegraphics[width=.99\textwidth]{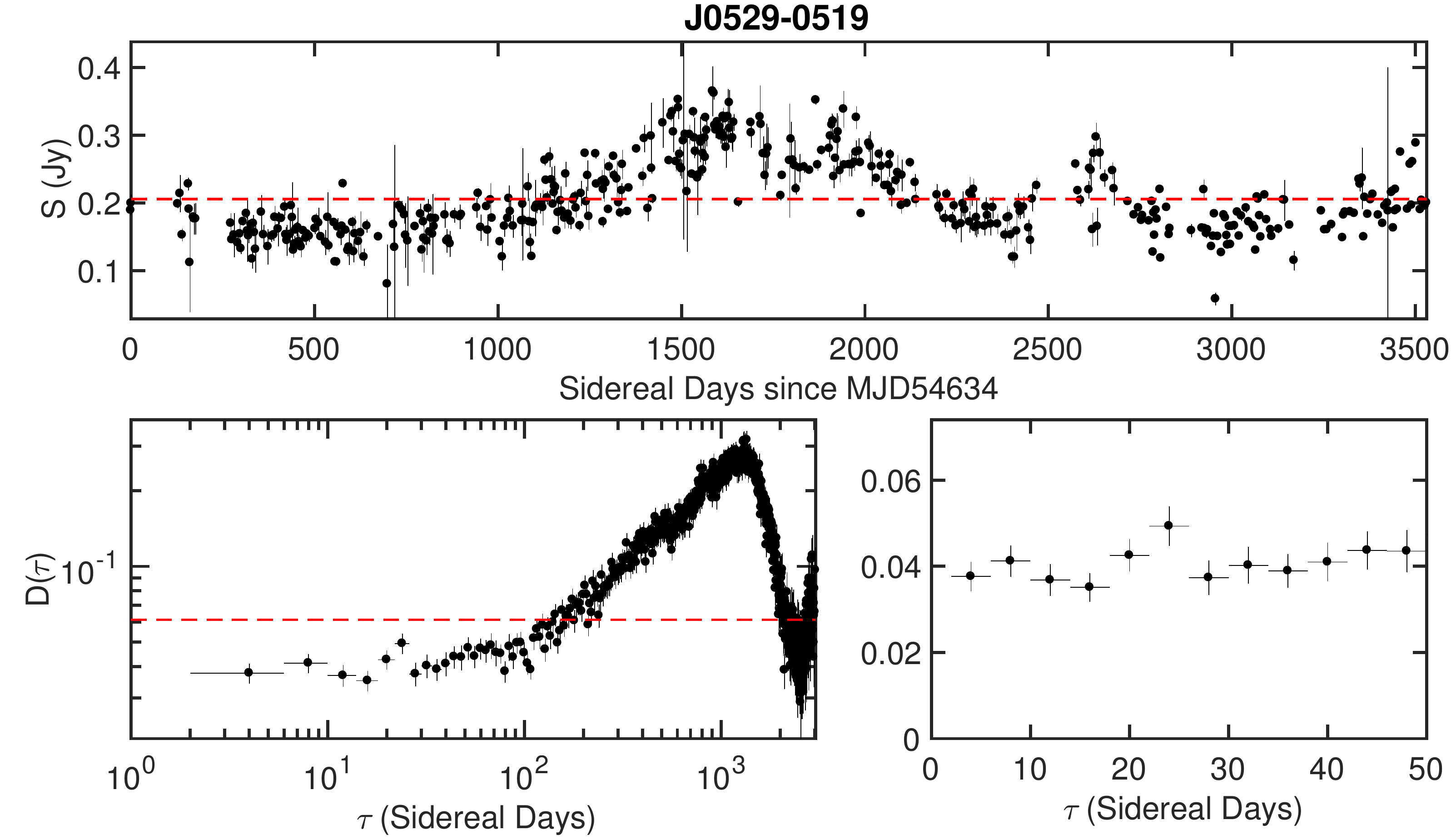}
		\caption{}
	\end{subfigure}%
	\begin{subfigure}[b]{.50\linewidth}
		\centering
		\includegraphics[width=.99\textwidth]{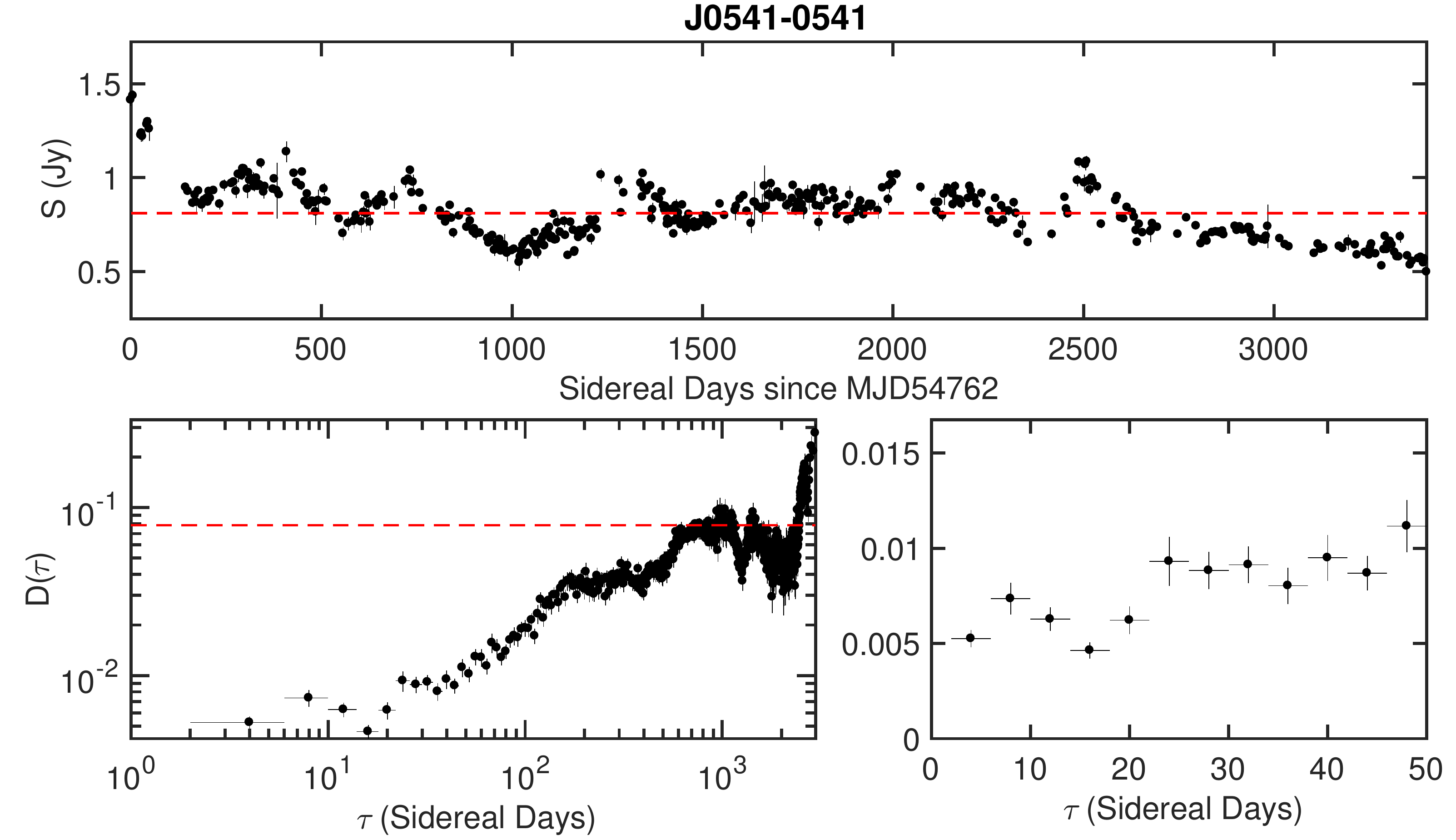}
		\caption{}
	\end{subfigure}\\%
	\begin{subfigure}[b]{.50\linewidth}
		\centering
		\includegraphics[width=.99\textwidth]{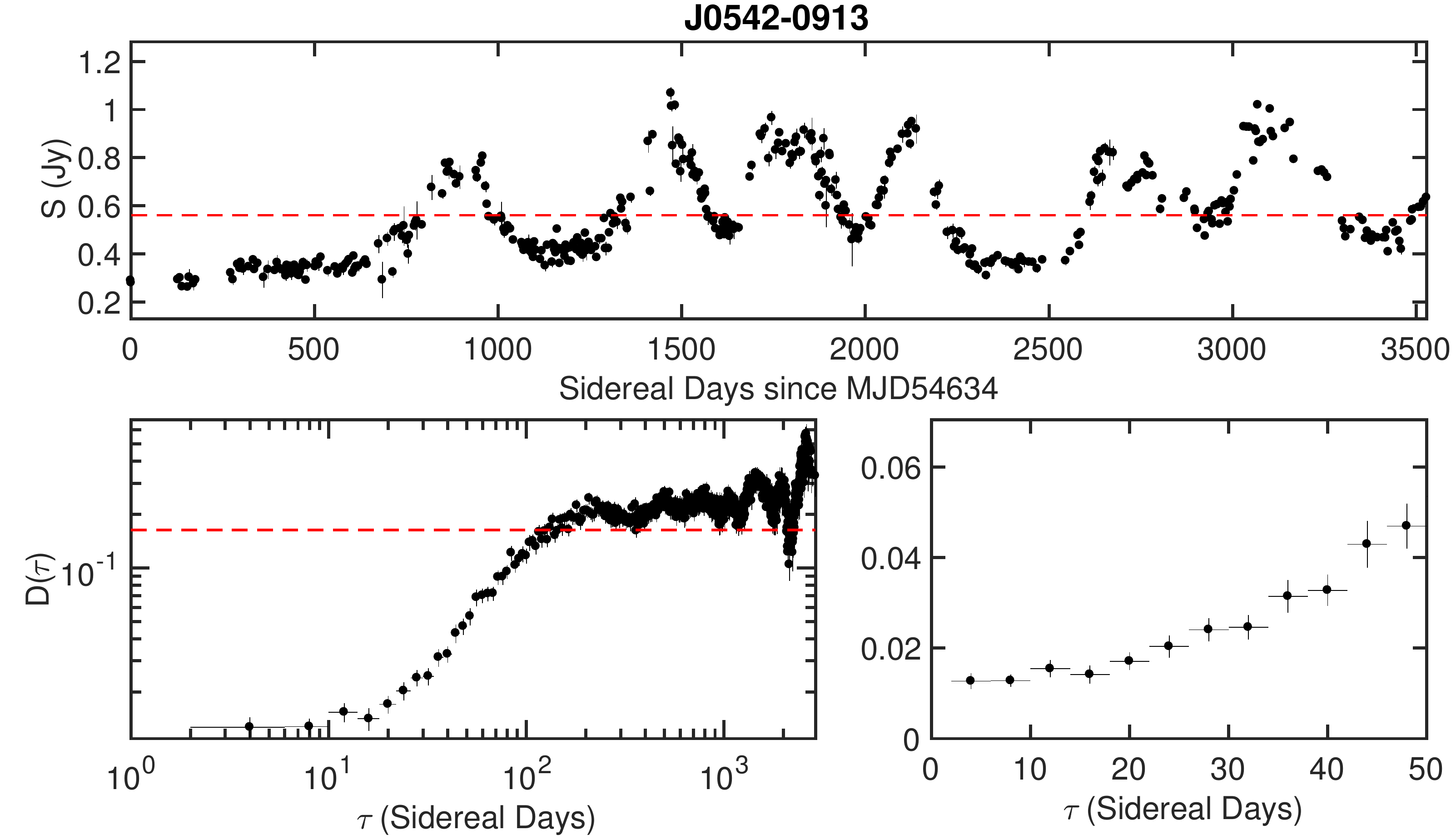}
		\caption{}
	\end{subfigure}%
    	\begin{subfigure}[b]{.50\linewidth}
    	\centering
    	\includegraphics[width=.99\textwidth]{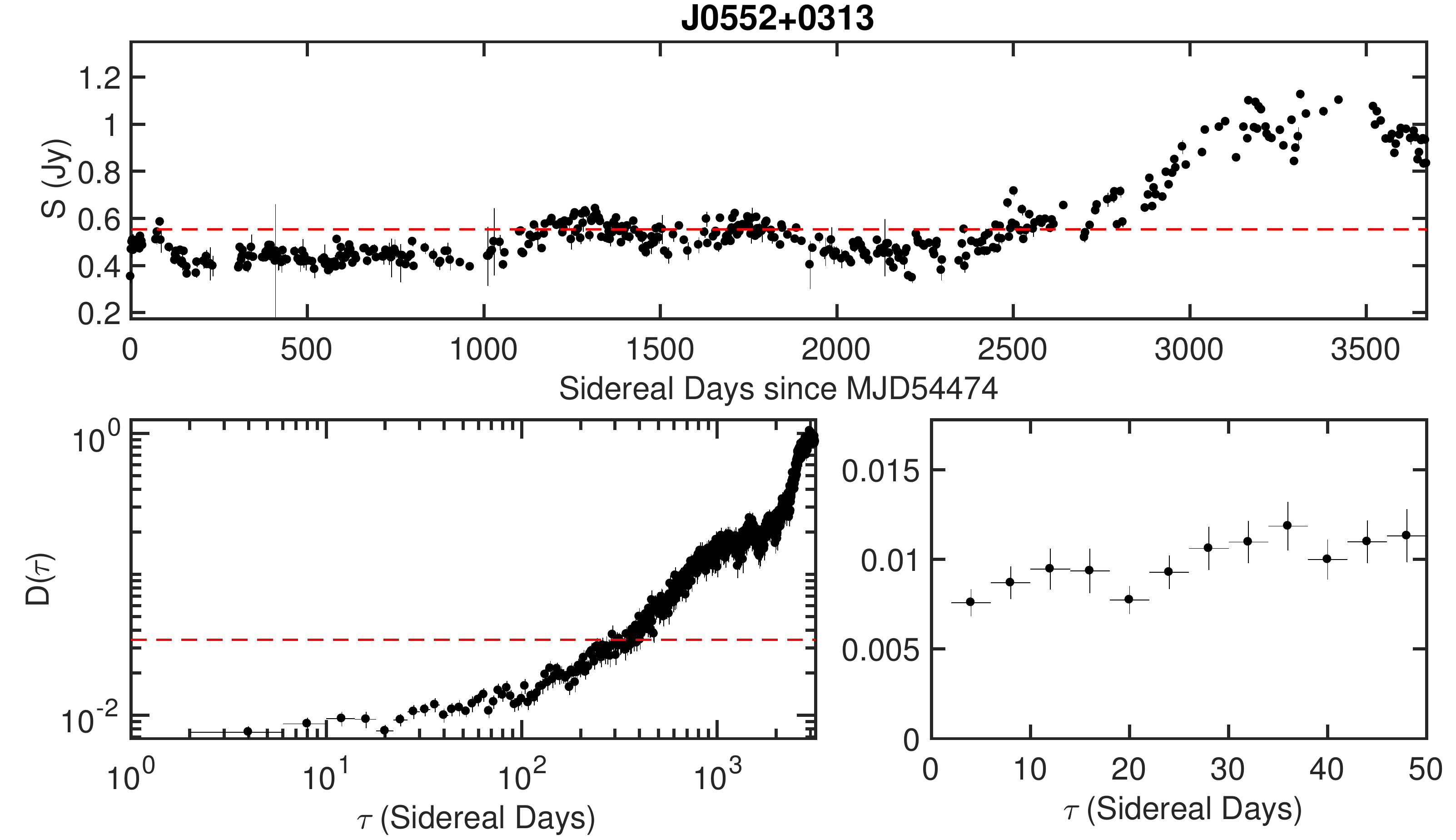}
    	\caption{}
    \end{subfigure}\\%
    \begin{subfigure}[b]{.50\linewidth}
    	\centering
    	\includegraphics[width=.99\textwidth]{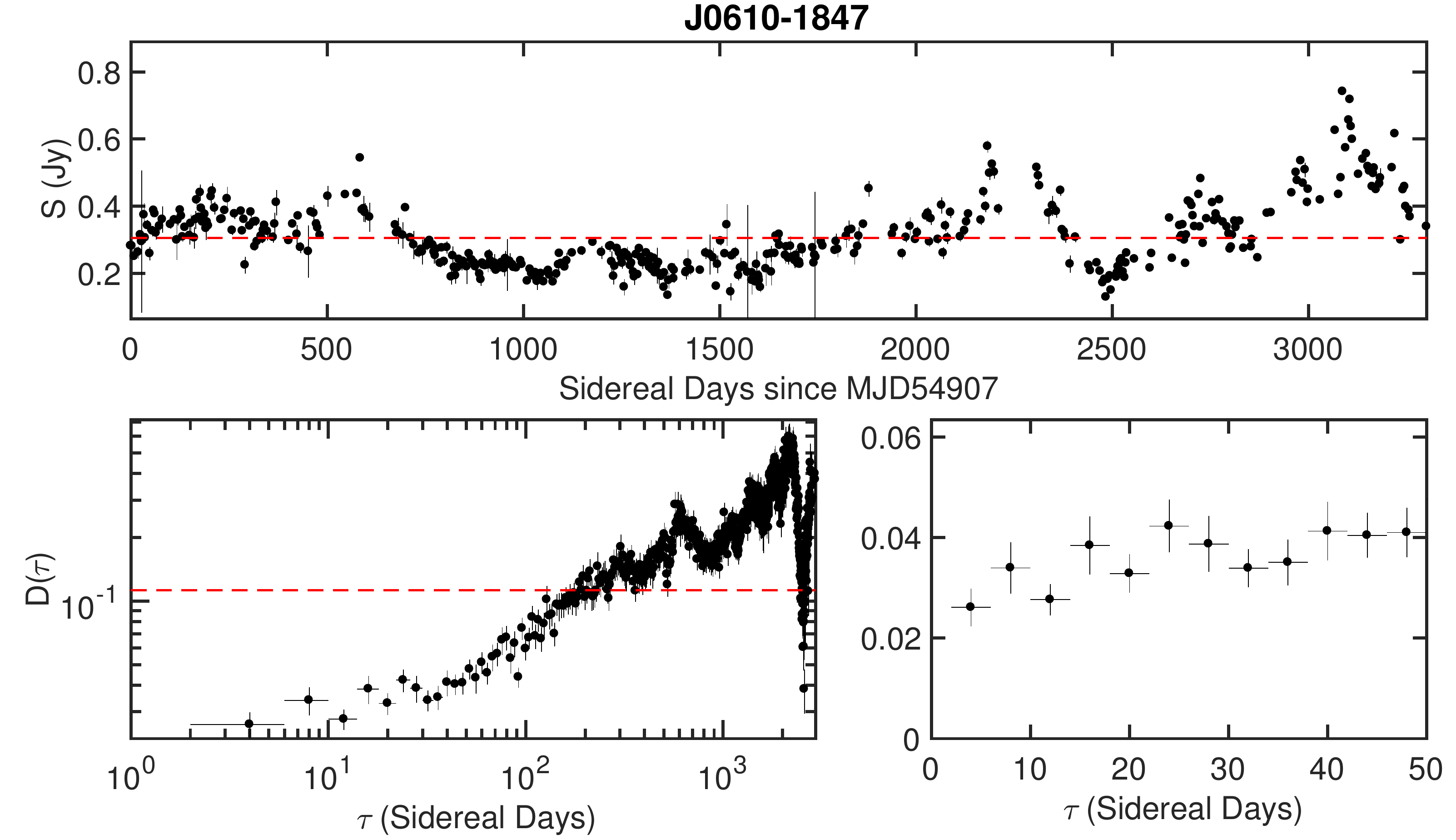}
    	\caption{}
    \end{subfigure}%
    \begin{subfigure}[b]{.50\linewidth}
    	\centering
    	\includegraphics[width=.99\textwidth]{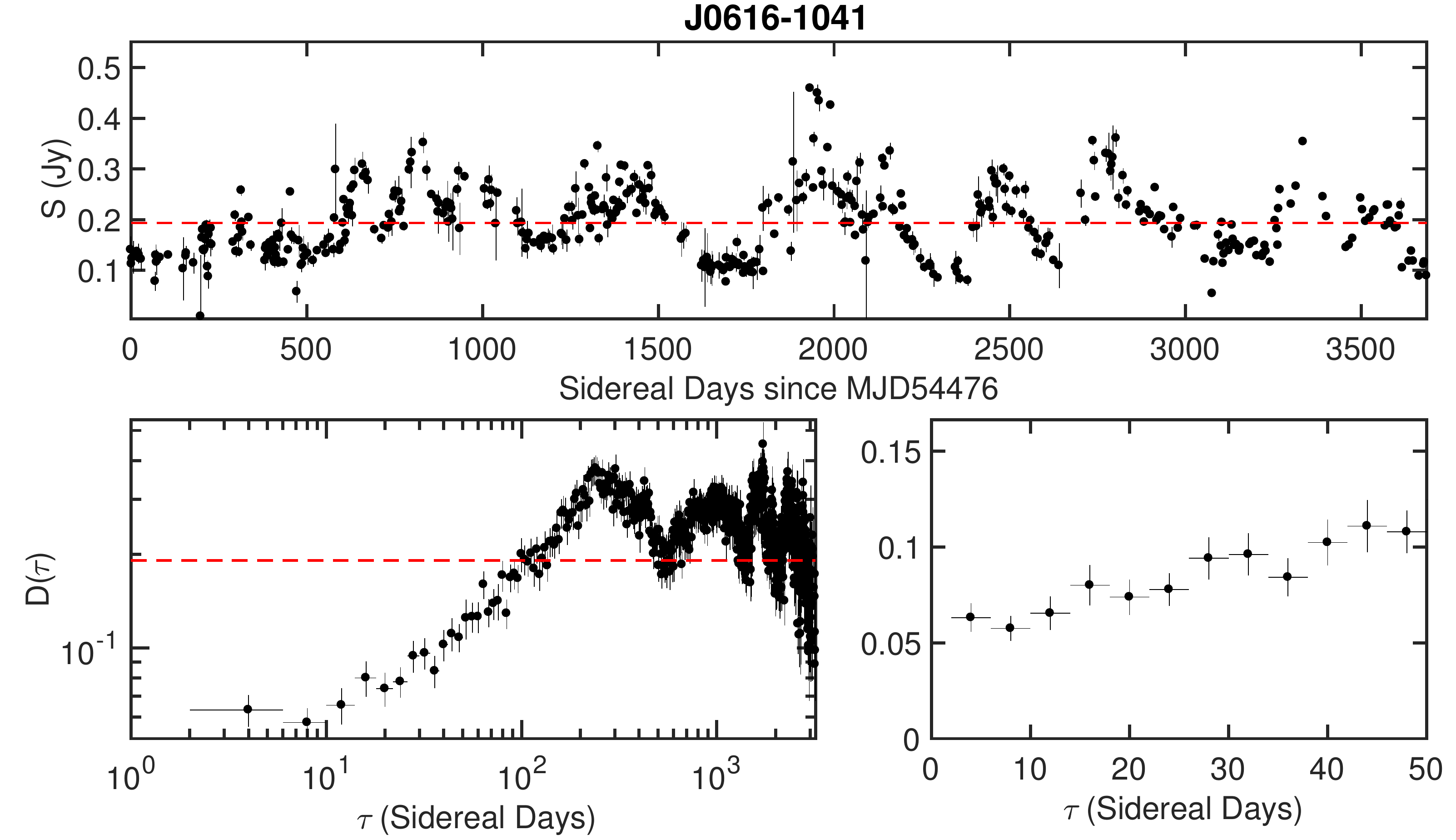}
    	\caption{}\label{J0616}
    \end{subfigure}\\%
    \begin{subfigure}[b]{.50\linewidth}
    	\centering
    	\includegraphics[width=.99\textwidth]{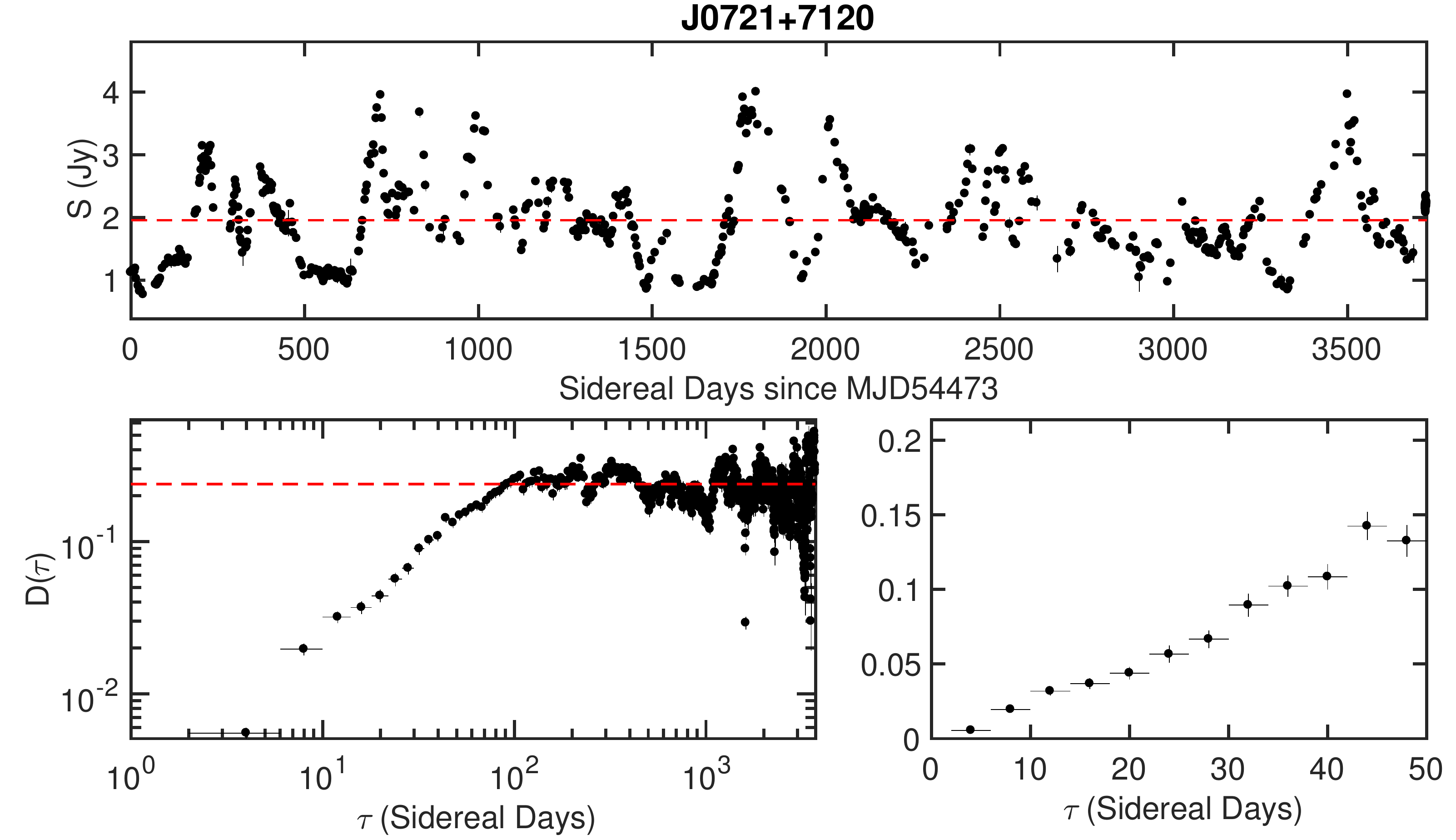}
    	\caption{}
    \end{subfigure}%
\begin{subfigure}[b]{.50\linewidth}
	\centering
	\includegraphics[width=.99\textwidth]{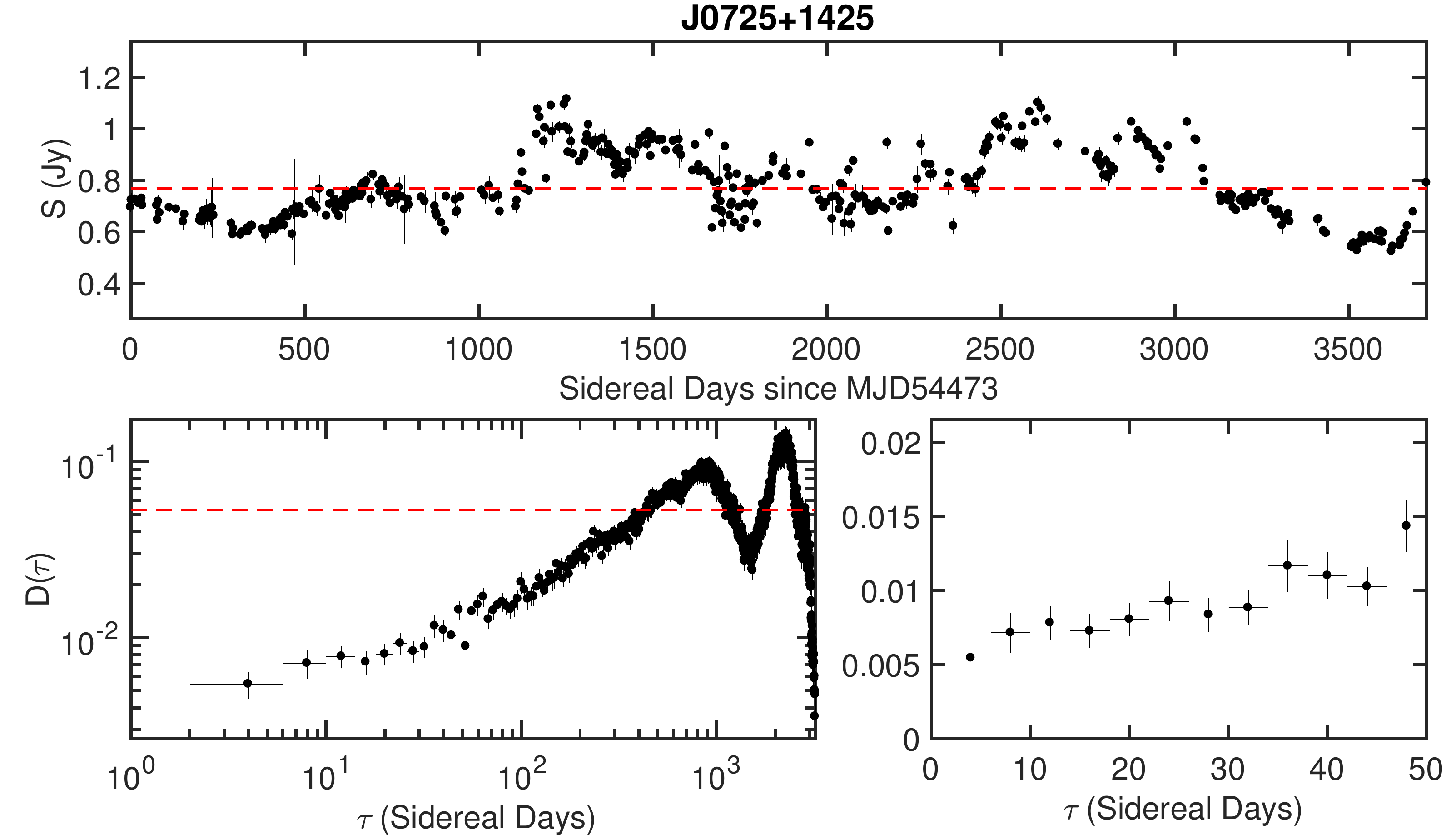}
	\caption{}
\end{subfigure}\\%
	\caption{Lightcurves and structure functions of the 20 sources exhibiting significant variability amplitudes on 4-day timescales, excluding J0259$-$0018. The lightcurves are shown in the top panel of each subfigure, where the horizontal dashed line denotes the mean flux density of the source. The error bars are given by Equation~\ref{erroreqrichards} \citep{richardsetal11}. The bottom panels of each subfigure show the structure function, $D(\tau)$, in its entirety in the left panel, and for $\tau \leq 50 \rm d$ in the right panel. The horizontal dashed line denotes $D_{m15}$ (Equation~\ref{m2D}) derived from the intrinsic modulation indices estimated by \citet{richardsetal14}.}
\end{figure*}

\begin{figure*}\ContinuedFloat
	\centering
	\begin{subfigure}[b]{.50\linewidth}
		\centering
		\includegraphics[width=.99\textwidth]{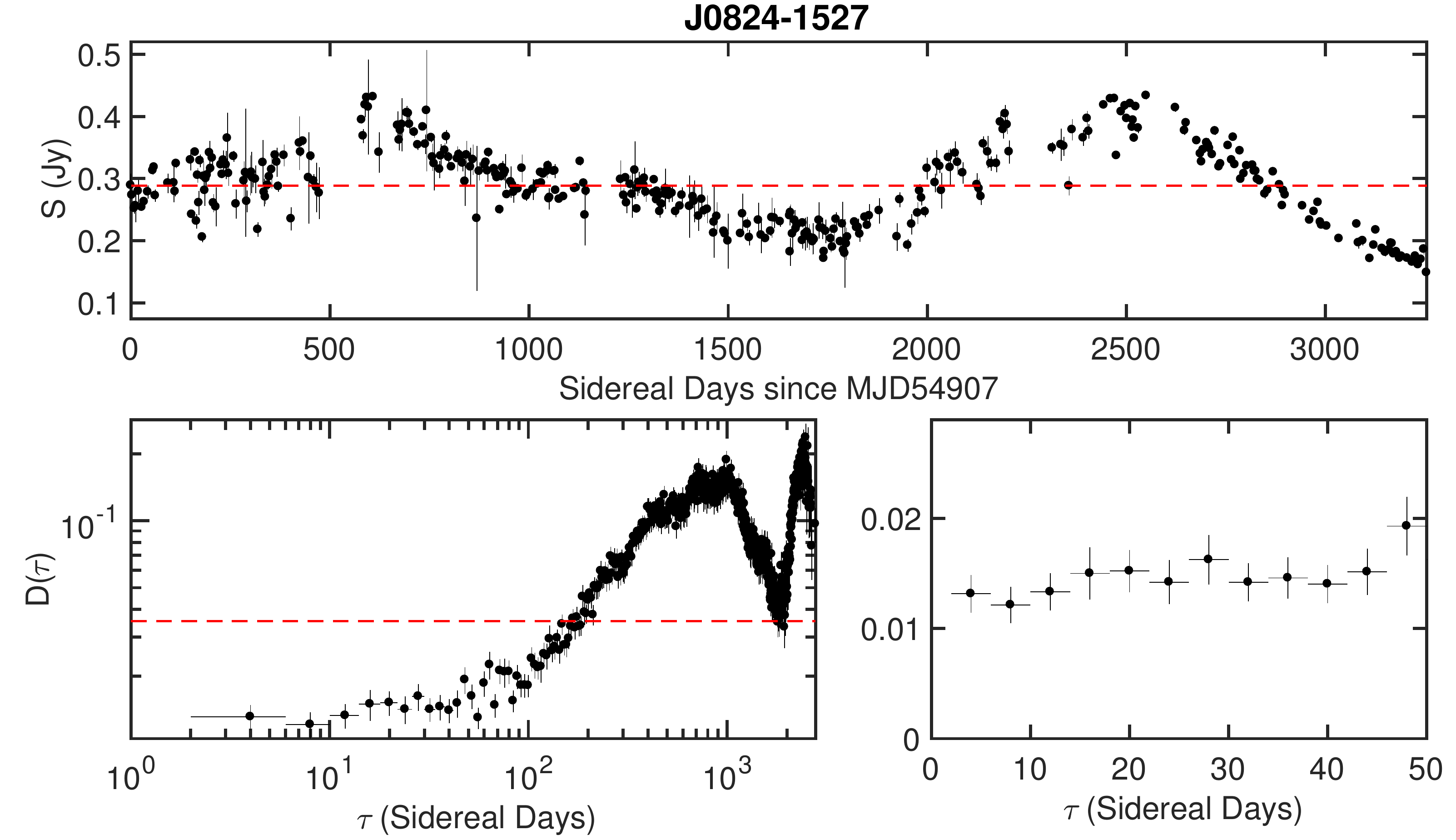}
		\caption{}
	\end{subfigure}%
	\begin{subfigure}[b]{.50\linewidth}
		\centering
		\includegraphics[width=.99\textwidth]{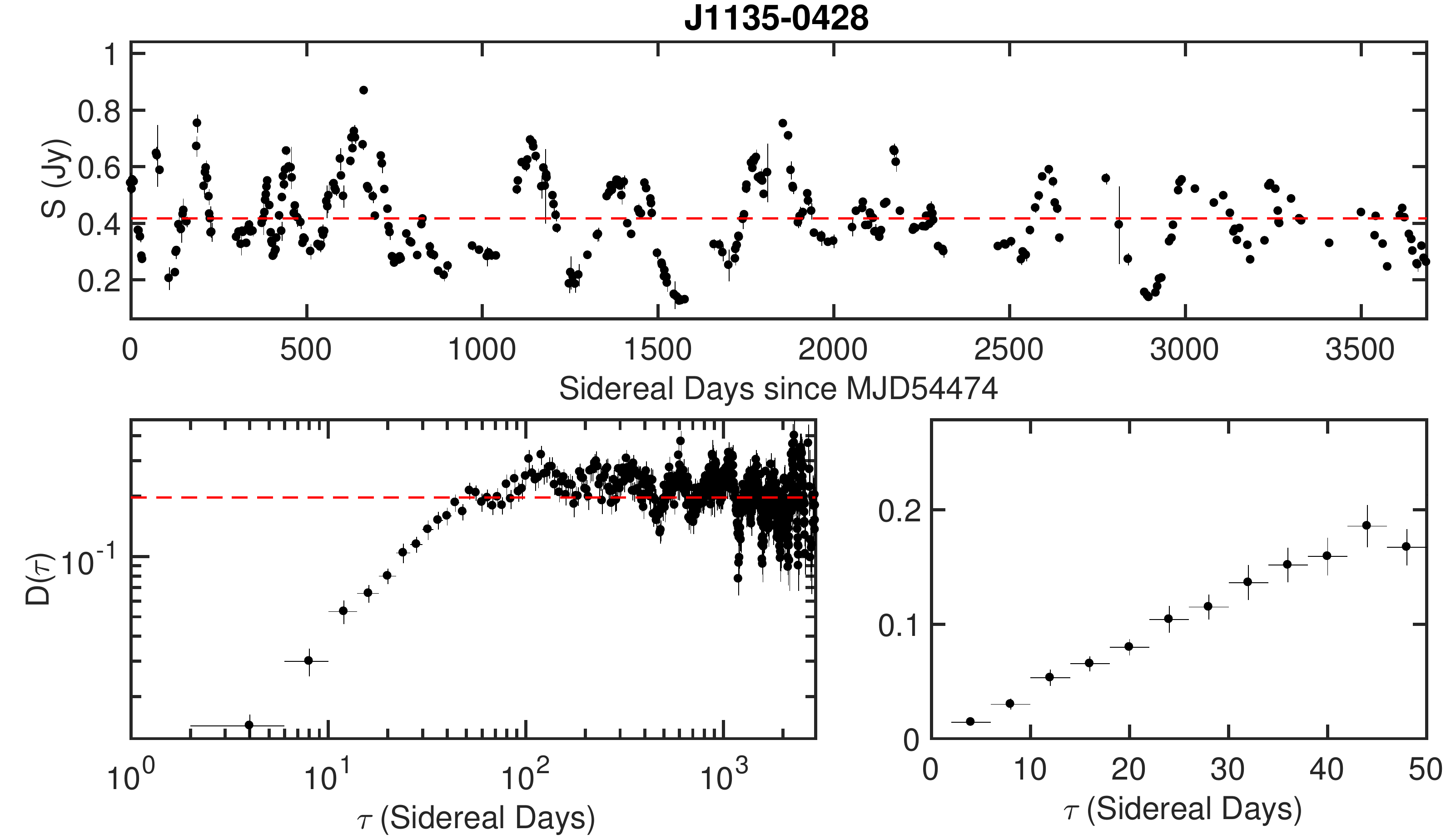}
		\caption{}
	\end{subfigure}\\%
	\begin{subfigure}[b]{.50\linewidth}
		\centering
		\includegraphics[width=.99\textwidth]{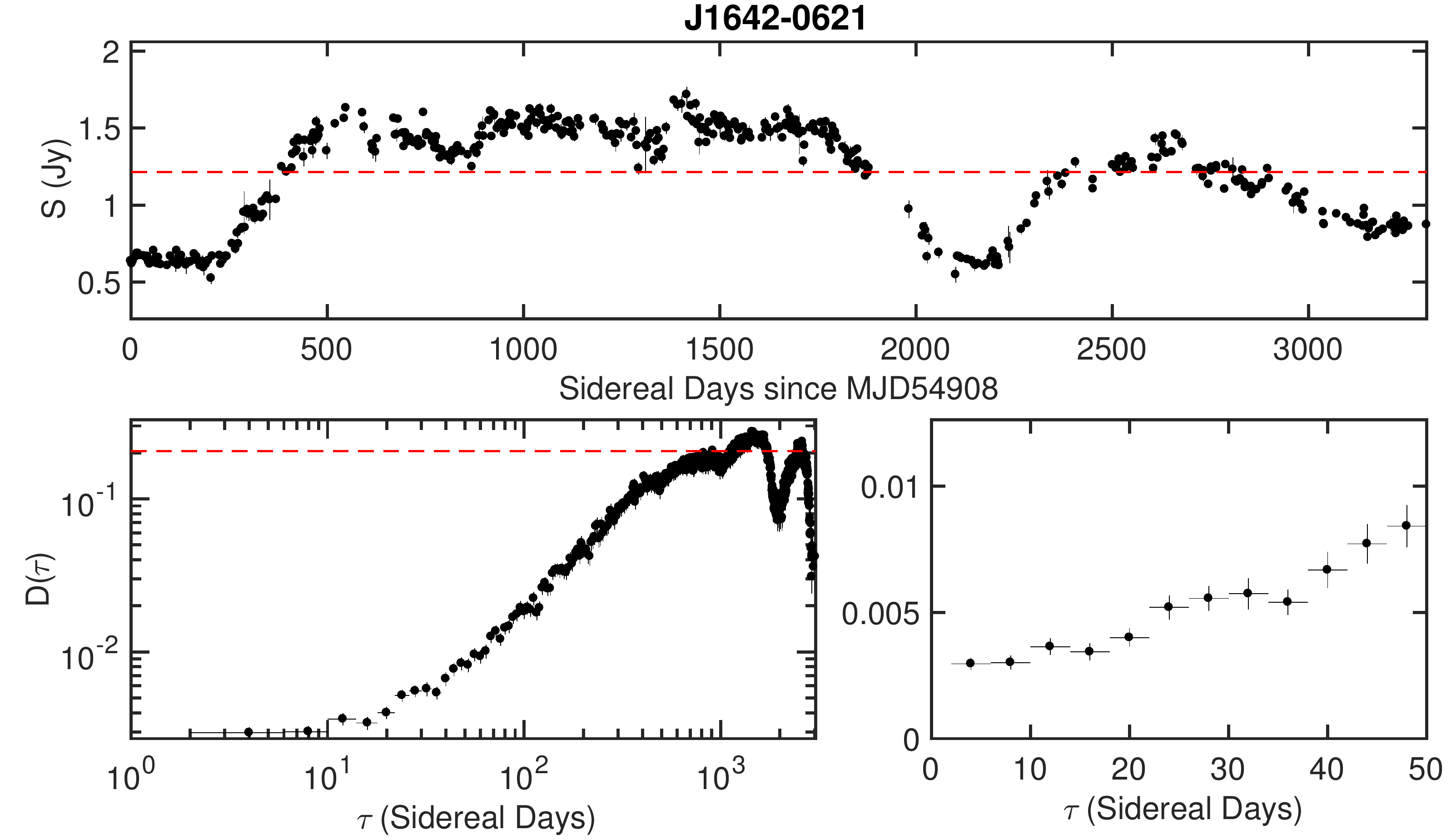}
		\caption{}
	\end{subfigure}%
	\begin{subfigure}[b]{.50\linewidth}
		\centering
		\includegraphics[width=.99\textwidth]{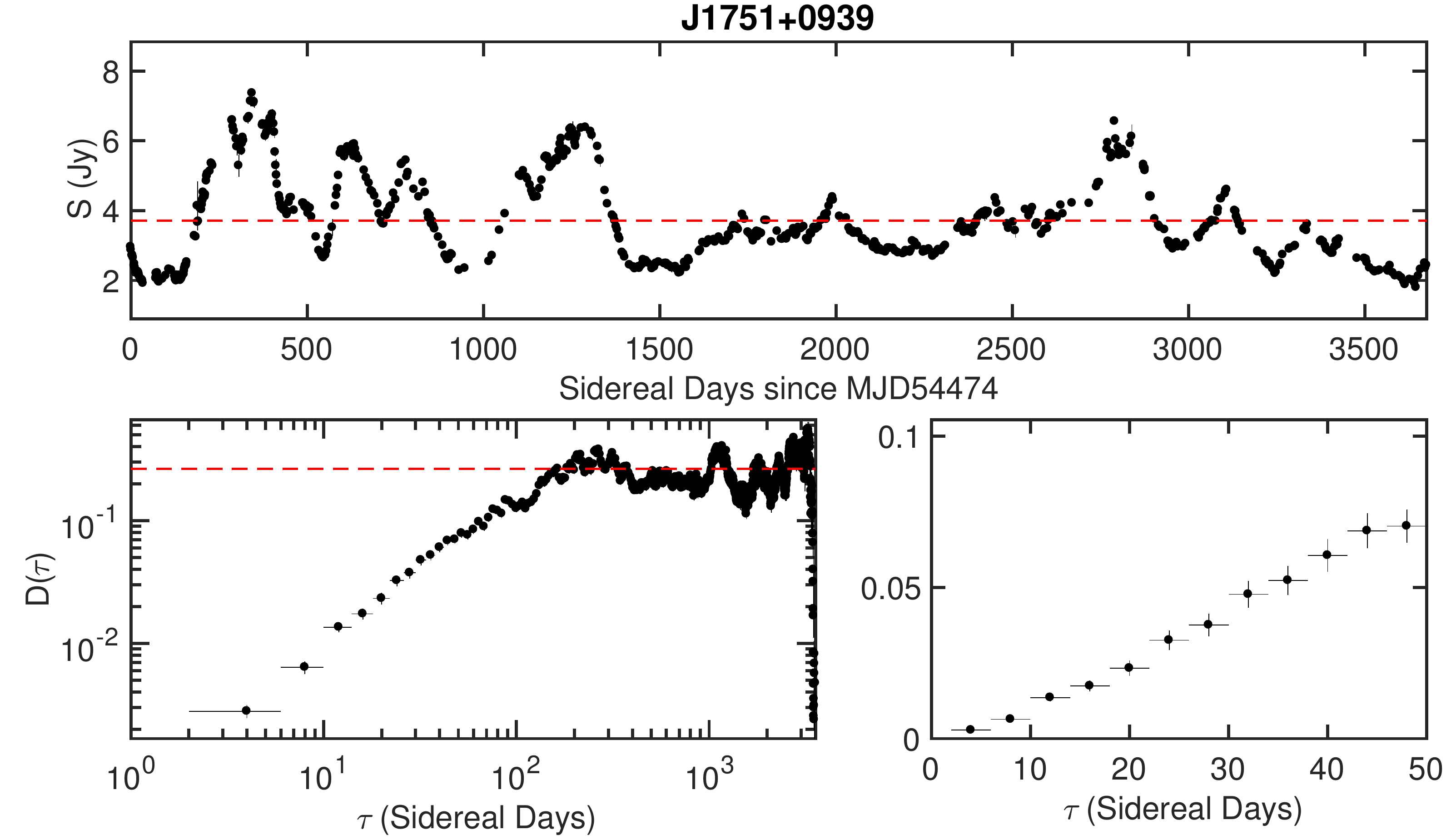}
		\caption{}
	\end{subfigure}\\%
 
	\caption{Lightcurves and structure functions of the 20 sources exhibiting significant variability amplitudes on 4-day timescales, excluding J0259$-$0018. The lightcurves are shown in the top panel of each subfigure, where the horizontal dashed line denotes the mean flux density of the source. The error bars are given by Equation~\ref{erroreqrichards} \citep{richardsetal11}. The bottom panels of each subfigure show the structure function, $D(\tau)$, in its entirety in the left panel, and for $\tau \leq 50 \rm d$ in the right panel. The horizontal dashed line denotes $D_{m15}$ (Equation~\ref{m2D}) derived from the intrinsic modulation indices estimated by \citet{richardsetal14}.}
\end{figure*}

\bsp	
\label{lastpage}
\end{document}